\documentstyle[psfig]{article}

\begin{document}

\title{Statistics of light in Raman and Brillouin nonlinear
couplers}

\author{J. Pe\v{r}ina, Jr. and J. Pe\v{r}ina \\
Joint Laboratory of Optics, Palack\'{y}
University and \\ Institute of Physics of Academy of
Sciences of the Czech Republic \\
and \\
Laboratory of Quantum Optics \\ Department of Optics,
Palack\'{y} University \\
17. listopadu 50, 772 07 Olomouc, Czech Republic}
\date{}

\maketitle

\begin{abstract}
Statistical properties of optical fields in nonlinear couplers
composed of two waveguides in which Raman
or Brillouin processes (with classical pumping)
are in operation and which are
mutually connected through the mutual Stokes and/or anti-Stokes linear
interactions are investigated in the framework of a generalized
superposition of coherent fields and quantum noise. Heisenberg
equations describing the couplers are solved both analytically
under special conditions and numerically in general cases.
Regimes for nonclassical properties of optical fields, such as
sub-Poissonian photon number statistics, negative reduced
moments of integrated intensity, and squeezing of quadrature
fluctuations are discussed in cases of single and compound fields.
General results are compared with those from the short-length approximation.
\end{abstract}

\section{Introduction}
Recently increasing attention has been devoted to the study of nonlinear
couplers composed of two or more waveguides with mutually
connected modes by means of evanescent waves \cite{1,3,mipe}.
Also quantum statistical
properties of such devices have been studied \cite{2,3}. Especially,
couplers composed of linear and
nonlinear waveguides have been examined
for both directional \cite{pp1,pp2} and
contradirectional \cite{pp2,pp3} propagation of light beams. Also
couplers composed of two nonlinear waveguides have been investigated
\cite{pp4}. Various nonlinear optical processes were assumed to operate
in nonlinear waveguides, such as second harmonic generation
\cite{pp4} (and references therein), frequency downconversion
\cite{jasi,mogkor,mogkope,korpe} and Kerr effect
\cite{cheba,korper,korolper}.

We examine quantum statistical properties of
couplers composed of two waveguides in which Raman
or Brillouin processes are active and in which modes (Stokes and
anti-Stokes) mutually linearly interact through evanescent waves.
We suppose classical strong laser pumping of the nonlinear processes
in both the waveguides. Then we can apply the method of the generalized
superposition of coherent fields and quantum noise to study spatial
dependence of
statistical quantities, such as photon number distribution,
reduced moments of integrated intensity, quadrature
variances and principal squeeze variance for both single
and compound modes. Special attention is devoted to nonclassical
properties of light exhibited by sub-Poissonian photon statistics,
negative reduced factorial moments or squeezing of quadrature
variances below the coherent-state values.
Statistical properties of modes involved in Raman
or Brillouin processes have been studied for a long time
(for reviews, see, e.g. \cite{ra,mi,pe}). Especially,
results contained in papers \cite{pie1,pie2,ka}, in which quantum
statistical properties of Brillouin \cite{pie1,pie2} and
Raman \cite{ka} processes were investigated in the framework
of the generalized superposition of coherent fields and quantum
noise, can be employed. The investigated model represents
a generalization of the above models in the direction
of incorporation of the influence of linear Stokes and
anti-Stokes interactions.

   Application of some earlier results for Raman scattering to
obtain sum and difference squeezing of vacuum fluctuations were
given in \cite{chiha,yeoha}, including the quantum correlations
between the Stokes and anti-Stokes fields, whereas the correlations
between the pump and Stokes waves were considered and measured
in \cite{vodko}. Raman effects can also play important role
in quantum soliton propagation \cite{we} and in solving various
inverse scattering problems \cite{geko}.

The dynamics of the couplers is solved analytically, numerically,
and in a short-length approximation in Section 2. Quantum
statistical properties of modes of the couplers are described
in Section 3. Section 4 contains a discussion of the statistical
behaviour of the systems under discussion based on results of the
numerical analysis. Conclusions are summarized in Section 5.

\section{Dynamics of nonlinear couplers}

A~model of the coupler composed of two mutually interacting waveguides
with Raman or Brillouin processes is quantally described by the following
momentum operator (for a scheme of involved interactions, see Fig. 1):
\begin{eqnarray}             
 \hat{G} &=& \sum_{j=L_1,S_1,A_1,V_1} \hbar k_j \hat{a}^\dagger_j
 \hat{a}_j + \left[ \hbar \tilde{g}_{A_1} \hat{a}_{L_1} \hat{a}_{V_1}
 \hat{a}^\dagger_{A_1} + \hbar \tilde{g}_{S_1} \hat{a}_{L_1}
 \hat{a}^\dagger_{V_1} \hat{a}^\dagger_{S_1} + \mbox{h.c.} \right]
 \nonumber \\
       & & \mbox{} + \sum_{j=L_2,S_2,A_2,V_2} \hbar k_j
 \hat{a}^\dagger_j
 \hat{a}_j + \left[ \hbar \tilde{g}_{A_2} \hat{a}_{L_2} \hat{a}_{V_2}
 \hat{a}^\dagger_{A_2} + \hbar \tilde{g}_{S_2} \hat{a}_{L_2}
 \hat{a}^\dagger_{V_2} \hat{a}^\dagger_{S_2} + \mbox{h.c.} \right]
 \nonumber \\
       & & \mbox{} + \left[ \hbar \tilde{\kappa}_S
 \hat{a}_{S_1}\hat{a}^\dagger_{S_2} + \hbar\tilde{\kappa}_A
 \hat{a}_{A_1} \hat{a}^{\dagger}_{A_2} + \mbox{h.c.} \right],
\end{eqnarray}
where $ \hat{a}_j $ ($ \hat{a}^\dagger_j $) are annihilation
(creation) operators of laser ($ L $), Stokes ($ S $),
anti-Stokes ($ A $), and vibration ($ V $) modes in the first
(1) and the second (2) waveguides. Wave vectors of the corresponding modes
are denoted as $ k_j $. Coefficients
$ \tilde{g}_{S_1} $ and $ \tilde{g}_{S_2} $ ($ \tilde{g}_{A_1} $
and $ \tilde{g}_{A_2} $) describe nonlinear Stokes (anti-Stokes)
interactions in the waveguides 1 and 2. Coefficients
$ \tilde{\kappa}_S $ and $ \tilde{\kappa}_A $ are related to linear
coupling between Stokes modes and anti-Stokes modes, respectively,
in both the waveguides
by means of the evanescent waves.
The symbol $ \hbar $ denotes the reduced Planck constant and
$ \mbox{h.c.} $ means Hermitian conjugate terms.
\begin{figure}        
  \centerline{\hbox{\psfig{file=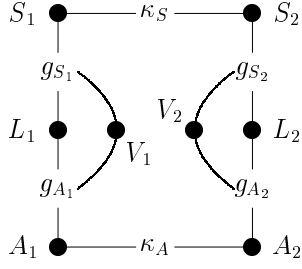,height=3.5cm}
     }}
  \vspace{3mm}
  \caption{Scheme of interactions in the quantum nonlinear
  coupler formed from two nonlinear
  waveguides operating by means of Brillouin or Raman processes, in which
  Stokes and anti-Stokes modes mutually interact through evanescent
  waves;
  $ S_j $, $ A_j $, $ L_j $, and $ V_j $ denote Stokes, anti-Stokes,
  laser pump, and phonon modes, $ g_{S_j} $
  ($ g_{A_j} $) is Stokes (anti-Stokes) nonlinear coupling
  constant in the $ j $th waveguide, and $ \kappa_S $ ($ \kappa_A $)
  represents Stokes (anti-Stokes) linear coupling constant.}
\end{figure}

The momentum operator $ \hat{G} $ in (1) reflets a symmetry
between the two waveguides, which is based on the exchange
of the suffices ($ L_1,S_1,A_1,V_1 $) and the suffices
($ L_2,S_2,A_2,V_2 $) with the simultaneous exchange of
constants $ \tilde{\kappa}_S $, $ \tilde{\kappa}_A $ with
their complex conjugate ones $ \tilde{\kappa}^*_S $,
$ \tilde{\kappa}^*_A $. This symmetry has to be conserved in
spatial development of operators and all quantum statistical
quantities, so we restrict our discussion of results only to modes
in waveguide 1; results valid for waveguide 2 are obtained simply
from the above mentioned symmetry.

Quantum spatial development of an operator $ \hat{a} $ is described
by the Heisenberg equation $ i\hbar \frac{d\hat{a}}{dz} =
[\hat{G},\hat{a}] $, where z~is a spatial variable along
the propagation direction, $ \hat{G} $
is the above introduced momentum operator and the symbol
$ [\;,\;] $ denotes a commutator. The Heisenberg equations
for the operators $ \hat{A}_j $ in the interaction picture
[$ \hat{A}_{j} =\hat{a}_j \exp(-ik_j z)$] and under the assumption
of strong classical laser pump
modes $ L_1 $ and $ L_2 $ ($ \hat{a}_j
\rightarrow \alpha_{j} \exp(ik_j z) $ for $ j= L_1,L_2 $,
$ \alpha_j $ being classical amplitudes) have the form
\begin{eqnarray}             
 \frac{d\hat{A}_{S_1}}{dz} &=& ig_{S_1}\hat{A}^{\dagger}_{V_1} + i\kappa^*_S\hat{A}_{S_2},
 \nonumber \\
 \frac{d\hat{A}_{A_1}}{dz} &=& ig_{A_1}\hat{A}_{V_1} + i\kappa^*_A\hat{A}_{A_2},
 \nonumber \\
 \frac{d\hat{A}_{V_1}}{dz} &=& ig^*_{A_1}\hat{A}_{A_1} +
 ig_{S_1}\hat{A}^{\dagger}_{S_1},
\end{eqnarray}
where
\begin{eqnarray}             
 g_{S_j} &=& \tilde{g}_{S_j} \alpha_{L_j} \exp(i\Delta k_{S_j}
 z),
  \nonumber \\
 g_{A_j} &=& \tilde{g}_{A_j} \alpha_{L_j} \exp(i\Delta k_{A_j}
 z),
  \mbox{\hspace{2cm} for $ j=1,2 $}, \nonumber \\
 \kappa_S &=& \tilde{\kappa}_S \exp(i\Delta K_S z), \nonumber \\
 \kappa_A &=& \tilde{\kappa}_A \exp(i\Delta K_A z)
\end{eqnarray}
with the phase mismatch vectors $ \Delta k_{S_j} =
k_{L_j} - k_{V_j} - k_{S_j} $, $ \Delta k_{A_j} =
k_{L_j} + k_{V_j} - k_{A_j} $ for $ j=1,2 $ and
$ \Delta K_S = k_{S_1} - k_{S_2} $, $ \Delta K_A =
k_{A_1} - k_{A_2} $.

The Heisenberg equations (2) are linear, which enables us to solve
them simply numerically and to retain operator character of the solution
(see the next section). They can be also solved analytically
under special conditions. Furthermore a short-length solution
can be reached. These three kinds of solutions will be described in the
following subsections.

The relation
\begin{equation}              
  \frac{d}{dz} \left[ \sum_{j=1,2} \left(
  \hat{A}^{\dagger}_{V_j}(z)\hat{A}_{V_j}(z) + \hat{A}^{\dagger}_{A_j}(z)\hat{A}_{A_j}(z)
  - \hat{A}^{\dagger}_{S_j}(z)\hat{A}_{S_j}(z) \right) \right] = 0,
\end{equation}
which represents a conservation law for photon numbers,
can be derived directly from (2).

\subsection{Numerical solution}

The set of equations (2) can be conveniently written in the
matrix form reflecting the symmetry of the model:
\begin{eqnarray}             
 \frac{d\hat{\bf A} }{dz} &=& i\hat{\bf M} \hat{\bf A}, \nonumber \\
  \hat{\bf A} &=& \left[ \begin{array}{c} \hat{\bf A}_1 \\
 \hat{\bf A}_2 \end{array} \right] ,
     \hat{\bf M} = \left[ \begin{array}{cc} \hat{\bf M}_1 &
     \hat{\bf M}_{12} \\ \hat{\bf M}_{12}^* &
     \hat{\bf M}_2 \end{array} \right] ,
\end{eqnarray}
where the vectors $ \hat{\bf A}_1 $, $ \hat{\bf A}_2 $ and the
matrices $ \hat{\bf M}_1 $, $ \hat{\bf M}_2 $, $ \hat{\bf M}_{12} $
are defined as follows
\begin{eqnarray}             
  \hat{\bf A}_j &=& \left[ \begin{array}{c} \hat{A}_{S_j} \\ \hat{A}^{\dagger}_{S_j} \\
  \hat{A}_{A_j} \\ \hat{A}^{\dagger}_{A_j} \\ \hat{A}_{V_j} \\ \hat{A}^{\dagger}_{V_j} \end{array}
  \right],
  \nonumber \\
  \hat{\bf M}_j &=& \left[ \begin{array}{cccccc}
   0 & 0 & 0 & 0 & 0 & g_{S_j} \\
   0 & 0 & 0 & 0 & -g_{S_j}^* & 0 \\
   0 & 0 & 0 & 0 & g_{A_j} & 0 \\
   0 & 0 & 0 & 0 & 0 & -g_{A_j}^* \\
   0 & g_{S_j} & g_{A_j}^* & 0 & 0 & 0 \\
   -g_{S_j}^* & 0 & 0 & -g_{A_j} & 0 & 0
   \end{array} \right]
  \mbox{\hspace{1cm} for $ j=1,2 $}, \nonumber \\
  \hat{\bf M}_{12} &=& \left[ \begin{array}{cccccc}
   \kappa_S^* & 0 & 0 & 0 & 0 & 0 \\
   0 & -\kappa_S & 0 & 0 & 0 & 0 \\
   0 & 0 & \kappa_A^* & 0 & 0 & 0 \\
   0 & 0 & 0 & -\kappa_A & 0 & 0 \\
   0 & 0 & 0 & 0 & 0 & 0 \\
   0 & 0 & 0 & 0 & 0 & 0
   \end{array} \right] .
\end{eqnarray}

The above set of equations is composed of two independent sets
of equations for operators ($ \hat{A}^{\dagger}_{S_1},\hat{A}_{A_1},\hat{A}_{V_1},
\hat{A}^{\dagger}_{S_2},\hat{A}_{A_2},\hat{A}_{V_2} $) and for their Hermitian
conjugates. It is more convenient to manipulate with only one
larger set of equations, because the procedure to solve
them can be easily generalized for the case of losses in
modes, in which also fluctuating Langevin forces must be taken
into account; this results in coupling of operators and their
conjugates through correlation functions of fluctuating
forces (see, e.g. \cite{pekre}).

Solution of equations (5) can be written in the form
$ \hat{\bf A}(z) = \exp(i\hat{\bf M} z) \hat{\bf A}(0) $ supposing
that the matrix $ \hat{\bf M} $ is $ z- $independent. Since we
assume in the following that all mismatches are zero, the matrix
$ \hat{\bf M} $ is really $ z- $independent.
Diagonalization of the matrix $ \hat{\bf M} $ then enables us to
determine spatial dependence of the evolution matrices
$ \hat{U} $ and $ \hat{V} $ in the operator solution (for
details, see e.g. \cite{pekre})
\begin{equation}             
 \hat{A}_{j}(z) = \sum_{k=1}^{6} \left[ U_{jk}(z) \hat{A}_{k}(0) +
 V_{jk}(z) \hat{A}^{\dagger}_{k}(0) \right] \mbox{\hspace{1cm} for
 $ j=1,\ldots,6 $},
\end{equation}
where the indices $ j,k $ run over the symbols $
S_1 $, $ A_1 $, $ V_1 $, $ S_2 $, $ A_2 $,
and $ V_2 $.

\subsection{Analytical solution}

In a special case, characterized by the conditions
$$                           
  |g_{S_1}| = |g_{S_2}|, |g_{A_1}| = |g_{A_2}|, |\kappa_S| =
  |\kappa_A|,  $$
\begin{equation}
  \frac{\kappa^*_A}{|\kappa_A|} \frac{g_{A_2}}{g_{A_1}} =
  -\frac{\kappa_S}{|\kappa_S|} \frac{g^*_{S_2}}{g^*_{S_1}},
\end{equation}
we can solve analytically the set of equations (5) using the
following transformations ($\kappa = |\kappa_S| $):
\begin{eqnarray}           
 \hat{b}_S &=& \frac{1}{\sqrt{2}} \exp(-i\kappa z) \left[
  \hat{A}_{S_1} + \frac{\kappa^*_S}{|\kappa_S|} \hat{A}_{S_2} \right] ,
  \nonumber \\
 \hat{c}_S &=& \frac{1}{\sqrt{2}} \exp(i\kappa z) \left[
  \hat{A}_{S_1} - \frac{\kappa^*_S}{|\kappa_S|} \hat{A}_{S_2} \right] ,
  \nonumber \\
 \hat{b}_A &=& \frac{1}{\sqrt{2}} \exp(-i\kappa z) \left[
  \hat{A}_{A_1} + \frac{\kappa^*_A}{|\kappa_A|} \hat{A}_{A_2} \right] ,
  \nonumber \\
 \hat{c}_A &=& \frac{1}{\sqrt{2}} \exp(i\kappa z) \left[
  \hat{A}_{A_1} - \frac{\kappa^*_A}{|\kappa_A|} \hat{A}_{A_2} \right] ,
  \nonumber \\
 \hat{b}_V &=& \frac{1}{\sqrt{2}} \exp(-i\kappa z) \left[
  \hat{A}_{V_1} + \frac{\kappa^*_A}{|\kappa_A|} \frac{g_{A_2}}{g_{A_1}}
  \hat{A}_{V_2} \right] ,
  \nonumber \\
 \hat{c}_V &=& \frac{1}{\sqrt{2}} \exp(i\kappa z) \left[
  \hat{A}_{V_1} - \frac{\kappa^*_A}{|\kappa_A|} \frac{g_{A_2}}{g_{A_1}}
  \hat{A}_{V_2} \right] .
\end{eqnarray}
The application of the transformations (9) results in the
following sets of equations:
\begin{eqnarray}             
 \frac{d}{dz} \left[ \begin{array}{c}
  \hat{b}_A \\ \hat{c}^\dagger_S \\ \hat{b}_V \end{array} \right]
  &=& i \left[ \begin{array}{ccc}
  0 & 0 & g_{A_1} \\ 0 & 0 & -g_{S_1}^* \\ g_{A_1}^* & g_{S_1} &
  - \kappa \end{array} \right]
  \left[ \begin{array}{c}
  \hat{b}_A \\ \hat{c}^\dagger_S \\ \hat{b}_V \end{array} \right] ,
  \nonumber \\
 \frac{d}{dz} \left[ \begin{array}{c}
  \hat{c}_A \\ \hat{b}^\dagger_S \\ \hat{c}_V \end{array} \right]
  &=& i \left[ \begin{array}{ccc}
  0 & 0 & g_{A_1} \\ 0 & 0 & -g_{S_1}^* \\ g_{A_1}^* & g_{S_1} &
  \kappa \end{array} \right]
  \left[ \begin{array}{c}
  \hat{c}_A \\ \hat{b}^\dagger_S \\ \hat{c}_V \end{array} \right]
\end{eqnarray}
and two sets obtained by Hermitian conjugation of the above equations.
These equations can be solved analytically (e.g. by finding
eigenvalues and eigenvectors) and the inverse transformation
leads to the solution of the operator equations (5) in the form:
\begin{eqnarray}             
 \hat{A}_{S_1}(z) &=& \frac{1}{r^2} \left[ -|g_{A_1}|^2 {\rm chh} +
  |g_{S_1}|^2 \left( -\frac{\kappa}{l^*} {\rm sh}\;{\rm sl}^* +
  {\rm ch}\;{\rm cl}^* \right) \right] \hat{A}_{S_1}(0) \nonumber \\
   & & \mbox{} +
  \frac{g_{A_1}g_{S_1}}{r^2} \left[ -{\rm chh} -
  \frac{\kappa}{l^*} {\rm sh}\;{\rm sl}^* + {\rm ch}\;{\rm cl}^*
  \right] \hat{A}^{\dagger}_{A_1}(0) \nonumber \\
  & & \mbox{} +
  i \frac{1}{r^2} \frac{\kappa_S^*}{|\kappa_S|} \left[
  -|g_{A_1}|^2{\rm shh} + |g_{S_1}|^2 \left( \frac{\kappa}{l^*}
  {\rm ch}\;{\rm sl}^* + {\rm sh}\;{\rm cl}^* \right) \right]
  \hat{A}_{S_2}(0) \nonumber \\
  & & \mbox{} +
  i \frac{g_{A_1}g_{S_1}}{r^2} \frac{\kappa_A}{|\kappa_A|}
  \left[ {\rm shh} - \frac{\kappa}{l^*} {\rm ch}\;{\rm sl}^* -
  {\rm sh}\;{\rm cl}^* \right] \hat{A}^{\dagger}_{A_2}(0) \nonumber \\
  & & \mbox{} +
  2i \frac{g_{S_1}}{l^*} {\rm ch}\;{\rm sl}^* \hat{A}^{\dagger}_{V_1}(0) -
  2 \frac{\kappa_S^*}{|\kappa_S|} \frac{g_{S_2}}{l^*}
  {\rm sh}\;{\rm sl}^* \hat{A}^{\dagger}_{V_2}(0),
 \nonumber \\
 \hat{A}_{A_1}(z) &=&
  \frac{g_{A_1}g_{S_1}}{r^2} \left[ {\rm chh} +
  \frac{\kappa}{l} {\rm sh}\;{\rm sl} - {\rm ch}\;{\rm cl}
  \right] \hat{A}^{\dagger}_{S_1}(0) +
  2i \frac{g_{A_1}}{l} {\rm ch}\;{\rm sl}\; \hat{A}_{V_1}(0) \nonumber \\
  & & \mbox{} +
  \frac{1}{r^2} \left[ |g_{S_1}|^2 {\rm chh} +
  |g_{A_1}|^2 \left( \frac{\kappa}{l} {\rm sh}\;{\rm sl} -
  {\rm ch}\;{\rm cl} \right) \right] \hat{A}_{A_1}(0) \nonumber \\
  & & \mbox{} +
  i \frac{1}{r^2} \frac{\kappa_A^*}{|\kappa_A|} \left[
  |g_{S_1}|^2 {\rm shh} -
  |g_{A_1}|^2 \left( \frac{\kappa}{l} {\rm ch}\;{\rm sl} +
  {\rm sh}\;{\rm cl} \right) \right] \hat{A}_{A_2}(0) \nonumber \\
  & & \mbox{} -
  i \frac{g_{A_1}g_{S_1}}{r^2} \frac{\kappa_S}{|\kappa_S|}\left[
  {\rm shh} - \frac{\kappa}{l} {\rm ch}\;{\rm sl} - {\rm sh}\;{\rm cl}
  \right] \hat{A}^{\dagger}_{S_2}(0) \nonumber \\
  & & \mbox{} -
  2 \frac{\kappa_A^*}{|\kappa_A|} \frac{g_{A_2}}{l} {\rm sh}\;{\rm sl}\;
  \hat{A}_{V_2}(0) , \nonumber \\
 \hat{A}_{V_1}(z) &=&
    2i \frac{g_{S_1}}{l} {\rm ch}\;{\rm sl}\; \hat{A}^{\dagger}_{S_1}(0) +
    2i \frac{g_{A_1}^*}{l} {\rm ch}\;{\rm sl}\; \hat{A}_{A_1}(0) +
    \left[ \frac{\kappa}{l} {\rm sh}\;{\rm sl} + {\rm ch}\;{\rm cl}
    \right] \hat{A}_{V_1}(0) \nonumber \\
    & & \mbox{} +
    2 \frac{g_{S_1}}{l} \frac{\kappa_S}{|\kappa_S|} {\rm sh}\;{\rm sl}\;
    \hat{A}^{\dagger}_{S_2}(0) -
    2 \frac{g_{A_1}^*}{l} \frac{\kappa_A^*}{|\kappa_A|}
    {\rm sh}\;{\rm sl}\; \hat{A}_{A_2}(0) \nonumber \\
    & & \mbox{} -
    i \frac{\kappa_A^*}{|\kappa_A|} \frac{g_{A_2}}{g_{A_1}}
    \left[ \frac{\kappa}{l} {\rm ch}\;{\rm sl} - {\rm sh}\;{\rm cl}
    \right] \hat{A}_{V_2}(0) .
\end{eqnarray}
The abbreviations
\begin{eqnarray}             
 {\rm shh} = \sin(\kappa z), & & {\rm chh} = \cos(\kappa z),
 \nonumber \\
 {\rm sh} = \sin(\kappa z/2), & & {\rm ch} = \cos(\kappa z/2),
 \nonumber \\
 {\rm sl} = \sin(l z/2), & & {\rm cl} = \cos(l z/2),
 \nonumber \\
  r = \sqrt{ |g_{S_1}|^2 - |g_{A_1}|^2 }, & &
  l = \sqrt{ \kappa^2 - 4r^2}
\end{eqnarray}
have been used in the above expressions. Expressions for
operators $ \hat{A}_{S_2} $, $ \hat{A}_{A_2} $, and $ \hat{A}_{V_2} $
are obtained from the expresions (11) using the before mentioned
symmetry of the coupler. Spatial development of creation
operators follows from Hermitian conjugation of the expressions
for annihilation operators.

\subsection{Short-length solution}

A~short-length solution of equations (5) valid up to
$ z^2 $ has the form:
\begin{eqnarray}             
 \hat{A}_{S_1}(z) &\approx& \hat{A}_{S_1} + i \left( g_{S_1} \hat{A}^{\dagger}_{V_1} +
 \kappa_S^* \hat{A}_{S_2} \right) z~\nonumber \\
 & & \mbox{} + \left( g_{S_1}g_{A_1}\hat{A}^{\dagger}_{A_1} + |g_{S_1}|^2
 \hat{A}_{S_1} - \kappa_S^*g_{S_2}\hat{A}^{\dagger}_{V_2} - |\kappa_S|^2 \hat{A}_{S_1}
 \right) z^2/2 \nonumber \\
 \hat{A}_{A_1}(z) &\approx& \hat{A}_{A_1} + i \left( g_{A_1} \hat{A}_{V_1} +
 \kappa_A^* \hat{A}_{A_2} \right) z~\nonumber \\
 & & \mbox{} - \left( g_{S_1}g_{A_1}\hat{A}^{\dagger}_{S_1} + |g_{A_1}|^2
 \hat{A}_{A_1} + \kappa_A^*g_{A_2}\hat{A}_{V_2} + |\kappa_A|^2 \hat{A}_{A_1}
 \right) z^2/2 \nonumber \\
 \hat{A}_{V_1}(z) &\approx& \hat{A}_{V_1} + i \left( g_{A_1}^* \hat{A}_{A_1} +
 g_{S_1}\hat{A}^{\dagger}_{S_1} \right) z~\nonumber \\
 & & \mbox{} - \left( (|g_{A_1}|^2 - |g_{S_1}|^2)
 \hat{A}_{V_1} + g_{A_1}^*\kappa_A^*
 \hat{A}_{A_2}  - g_{S_1}\kappa_S \hat{A}^{\dagger}_{S_2}
 \right) z^2/2 .
\end{eqnarray}
Operators on the right hand sides of expressions in (13) are taken at
$ z=0 $. Expressions for the other
operators are reached from the symmetry and by means of Hermitian
conjugation of the above expressions.

\section{Quantum statistical properties of nonlinear couplers}

Quantum statistical properties of optical fields involved in the above
processes can be studied
in the framework of the generalized superposition of coherent
fields and quantum noise (\cite{pe}, Secs. 8.5, 9.3, and 9.4),
which is characterized by the normal
characteristic function $ C_{\cal N}(\{\beta_j\},z) $
in the Gaussian form:
\begin{eqnarray}             
C_{\cal N}(\{\beta_j\},z) & = & \exp \left\{ \sum_{j=1}^{6}
\left[-B_j(z)
|\beta_j|^2 + \left(\frac{1}{2} C_j(z) \beta_j^{*2} + \mbox{c.c.}\right)
\right] \right. \nonumber\\
& &  \mbox{} + \sum_{j=1}^{6} \sum_{k=1, j<k}^{6}[ D_{jk}(z)
\beta_j^* \beta_k^* + \bar{D}_{jk}(z)
\beta_j \beta_k^* + \mbox{c.c.}]     \nonumber\\
& & \left. \mbox{} + \sum_{j=1}^{6} [ \beta_j \xi_j^*(z)
- \mbox{c.c.}]\right\},
\end{eqnarray}
where c.c. denotes the complex conjugate terms. The complex
amplitudes $ \xi_1(z) $, $ \ldots $, $ \xi_6(z) $
correspond to the annihilation
operators $ \hat{A}_{S_1}(z) $, $ \hat{A}_{A_1}(z) $, $ \hat{A}_{V_1}(z) $,
$ \hat{A}_{S_2}(z) $,
$ \hat{A}_{A_2}(z) $, and $ \hat{A}_{V_2}(z) $, respectively.
The quantum noise functions $ B_j(z) $, $ C_j(z) $,
$ D_{jk}(z) $, and $ \bar{D}_{jk}(z) $ defined by the relations
\begin{eqnarray}             
 B_j(z) &=& \langle \Delta \hat{A}^{\dagger}_{j}(z) \Delta \hat{A}_{j}(z) \rangle,
  \nonumber \\
 C_j(z) &=& \langle (\Delta \hat{A}_{j}(z))^2 \rangle,
  \nonumber \\
 D_{jk}(z) &=& \langle \Delta \hat{A}_{j}(z) \Delta \hat{A}_{k}(z) \rangle,
  \nonumber \\
 \bar{D}_{jk}(z) &=& - \langle \Delta \hat{A}^{\dagger}_{j}(z) \Delta \hat{A}_{k}(z) \rangle,
\end{eqnarray}
can be determined in terms of values of these quantities for incident
fields, $ B_j $, $ C_j $
($ B_j = B_j(0) + 1 $, $ C_j = C_j(0) $ if antinormal ordering of field
operators is adopted),
provided that the operator solution of the corresponding Heisenberg
equations is known (for details, see e.g. \cite{pj}).
Antinormal ordering adopted for incident field operators
enables us to describe nonclassical states of light. Values
of quantities $ B_j $ and $ C_j $ are given in the case of the
squeezed fields with additional noise by
\begin{eqnarray}       
B_j &=& \cosh^2 (r_j)  +  \langle n_{chj} \rangle,  \nonumber\\
C_j &=& \frac{1}{2} \exp(i \theta_j) \sinh(2 r_j) ,
\end{eqnarray}
where $ r_j $ are the squeeze parameters of the incident beams,
$ \theta_j $ are the squeeze phases and
$ \langle n_{chj} \rangle $ represent the mean number of
external noise particles in the $ j $th mode.
In general signal squeezed light with an additional
noise is described by the above expressions; when
$ r_j = 0 $, $ \langle n_{chj}\rangle = 0 $ ($ B_j = 1 $,
$ C_j = 0 $) we have a coherent state and the
superposition of the signal coherent field and noise is characterized
by $ r_j =0 $, $ \langle n_{chj}
\rangle \neq 0 $ ($ B_j = \langle n_{chj}\rangle + 1 $, $ C_j = 0 $).

The photon number distribution $ p(n_j,z) $ and the
moments of the integrated intensity $ \langle W^k_j (z)\rangle $
can be obtained from the normal characteristic function
$ C_{\cal N}({\beta_j},z) $ given in (14) in terms of the Laguerre
polynomials for both single and compound (two-mode)
fields (for details, see e.g. \cite{pp4,pe,pj} and references
therein). Expressions for variances of quadrature components
$ \langle (\Delta \hat{p}_j(z))^2 \rangle $,
$ \langle (\Delta \hat{q}_j(z))^2 \rangle $,
for principal squeeze variance $ \lambda_j(z) $, and for
uncertainty product $ u_j(z) $ can also be found in
\cite{pp4,pe,pj}.

We introduce only expressions for the variances
$ \langle (\Delta W)^2 \rangle $ of the integrated intensity
and for the principal squeeze variances $ \lambda $ for single
and compound modes, which will be used when analyzing the statistical
properties on the basis of the short-length solution.

We get the variance
$ \langle (\Delta W_j)^2 \rangle $ of the integrated intensity
in the form
(\cite{pe}, p. 268)
\begin{equation}             
 \langle [\Delta W_j(z)]^2 \rangle = B_j^2(z) + |C_j(z)|^2 +
 2 B_j(z)|\xi_j(z)|^2 + \left( C_j(z) \xi_j^{*2}(z) +
 \mbox{c.c.} \right)
\end{equation}
for the $ j $th single mode. The variance of the integrated
intensity of the ($ j,k $)th compound mode equals
\begin{equation}             
 \langle [\Delta W_{jk}(z)]^2 \rangle = \langle [\Delta W_j(z)]^2 \rangle
 + \langle [\Delta W_k(z)]^2 \rangle + 2 \langle \Delta W_j(z)
 \Delta W_k(z) \rangle   ,
\end{equation}
where the correlation of fluctuations of integrated intensities
is given by
\begin{eqnarray}             
 \langle \Delta W_j(z) \Delta W_k(z) \rangle &=& |D_{jk}(z)|^2 +
 |\bar{D}_{jk}(z)|^2  \nonumber \\
 & & \mbox{} + \left( D_{jk}(z) \xi_j^*(z) \xi_k^*(z)
 - \bar{D}_{jk}(z) \xi_j(z) \xi_k^*(z) + \mbox{c.c.} \right).
\end{eqnarray}

The principal squeeze variances $ \lambda_j(z) $ for the $ j $th single mode
and $ \lambda_{jk}(z) $ for the ($ j,k $)th compound mode
are expressed as \cite{pp4}:
\begin{eqnarray}            
 \lambda_j(z) &=& 1 + 2[ B_j(z) - |C_j(z)| ] , \nonumber \\
 \lambda_{jk}(z) &=& 2 \{ 1 + B_j(z) + B_k(z) -
 (\bar{D}_{jk}(z) + \mbox{c.c.} ) \nonumber \\
 & & \mbox{} - |C_j(z) + C_k(z) +
 2 D_{jk}(z)| \} .
\end{eqnarray}

We now can pay our attention to the statistical properties
of the short-length solution. We firstly discuss
the case with all modes being at the beginning
of the interaction in coherent states (Brillouin
processes). The coefficients of the normal characteristic function
$ C_{\cal N}(\{\beta_j\},z) $ read up to $ z^2 $:
\begin{eqnarray}             
 B_{S_1}(z) &=& |g_{S_1}|^2 z^2 , \nonumber \\
 B_{V_1}(z) &=& |g_{S_1}|^2 z^2 , \nonumber \\
 D_{S_1A_1}(z) &=& -g_{A_1}g_{S_1} z^2/2 , \nonumber \\
 D_{S_1V_1}(z) &=& ig_{S_1}z , \nonumber \\
 D_{S_1V_2}(z) &=& - g_{S_2}\kappa_S^* z^2 /2 .
\end{eqnarray}
The other coefficients (not obtained from the symmetry)
are zero. We get from (17) and (21) for single mode variances of
the integrated intensity
\begin{eqnarray}            
 \langle (\Delta W_{S_1}(z))^2 \rangle &=& 2 |g_{S_1}|^2
 |\xi_{S_1}|^2 z^2 , \nonumber \\
 \langle (\Delta W_{A_1}(z))^2 \rangle &=& 0 , \nonumber \\
 \langle (\Delta W_{V_1}(z))^2 \rangle &=& 2 |g_{S_1}|^2
 |\xi_{V_1}|^2 z^2 .
\end{eqnarray}
The symbols $ \xi $ denote initial coherent amplitudes
of the radiation and phonon modes.
The variances of the integrated intensity for compound modes,
which can be negative, read
\begin{eqnarray}           
 \langle (\Delta W_{S_1A_1}(z))^2 \rangle &=& 2 |g_{S_1}|^2
 |\xi_{S_1}|^2 z^2  - ( g_{A_1}g_{S_1} \xi_{S_1}^* \xi_{A_1}^*
  z^2 + \mbox{c.c.} ), \nonumber \\
 \langle (\Delta W_{S_1V_1}(z))^2 \rangle &=& 2 [
 (ig_{S_1}\xi_{S_1}^* \xi_{V_1}^*z + \mbox{c.c.}) +
 |g_{S_1}|^2 (1 + 3|\xi_{S_1}|^2 + 3|\xi_{V_1}|^2) z^2 \nonumber \\
 & & \mbox{} +
 (g_{S_1}g_{A_1} \xi_{A_1}^* \xi_{S_1}^* z^2 + \mbox{c.c.} ) +
 (g_{S_1} \kappa_S \xi_{S_2}^* \xi_{V_1}^* z^2  + \mbox{c.c.}) ],
 \nonumber \\
 \langle (\Delta W_{S_1V_2}(z))^2 \rangle &=& 2 |g_{S_1}|^2
 |\xi_{S_1}|^2 z^2 + 2|g_{S_2}|^2 |\xi_{V_2}|^2 z^2
 - ( g_{S_2} \kappa_S^* \xi_{S_1}^* \xi_{V_2}^*
  z^2 + \mbox{c.c.} ). \nonumber \\
\end{eqnarray}
From the point of view of the short-length solution,
negative variances of the integrated intensity reflecting
nonclassical properties of light can be reached
only in the compound modes ($ S_1,A_1 $), ($ S_1,V_1 $), and
($ S_1,V_2 $) when initial phases are suitably chosen.
The linear coupling between Stokes modes (described by $ \kappa_S $)
leads to the negative intergrated intensity variance
in the compound mode
($ S_1,V_2 $) and it also can support the negative integrated
intensity variance of
the mode ($ S_1,V_1 $).

The coefficients $ C_j(z) $ are zero not only in the
short-length solution but also when we suppose
initial non-squeezed light ($ C_j(0) = 0 $). This stemms
from the momentum operator (1), which contains only first
powers of operators. This excludes squeezing in single mode
fields. Squeezing of vacuum fluctuations can be reached
in the compound modes ($ S_1,A_1 $), ($ S_1,V_1 $), and
($ S_1,V_2 $), the principal squeeze variances of which are given by,
using (20) and (21):
\begin{eqnarray}            
 \lambda_{S_1A_1}(z) &=& 2[ 1 + |g_{S_1}| ( |g_{S_1}| - |g_{A_1}|)
  z^2 ] , \nonumber \\
 \lambda_{S_1V_1}(z) &=& 2[ 1 - 2|g_{S_1}| z~+ |g_{S_1}|^2 z^2 ]
 , \nonumber \\
  \lambda_{S_1V_2}(z) &=& 2[ 1 + ( |g_{S_1}|^2 + |g_{S_2}|^2 ) z^2
  - |g_{S_2}| |\kappa_S| z^2 ].
\end{eqnarray}
The linear Stokes coupling $ \kappa_S $ thus can lead to
generation of squeezed light in the mode ($ S_1,V_2 $).

Now using the short-length solution we analyze statistical properties of
modes for Raman processes, i.e. when phonons are initially chaotic and
the other modes are initially coherent. Coefficients in the
normal characteristic function $ C_{\cal N}(\{\beta_j\},z) $
have the form:
\begin{eqnarray}           
 B_{S_1}(z) &=& |g_{S_1}|^2 (n_{V_1} + 1 ) z^2 ,\nonumber \\
 B_{A_1}(z) &=& |g_{A_1}|^2 n_{V_1} z^2 , \nonumber \\
 B_{V_1}(z) &=& n_{V_1} + |g_{S_1}|^2 (n_{V_1} + 1 ) z^2 -
 |g_{A_1}|^2 n_{V_1} z^2 ,\nonumber \\
 D_{S_1A_1}(z) &=& - g_{S_1} g_{A_1} (n_{V_1} + 1/2 ) z^2 , \nonumber \\
 D_{S_1V_1}(z) &=& i g_{S_1} (n_{V_1} + 1 ) z~, \nonumber \\
 D_{S_1V_2}(z) &=& - g_{S_2}\kappa_S^*(n_{V_2} + 1) z^2 /2, \nonumber \\
 \bar{D}_{A_1V_1}(z) &=& i g_{A_1}^* n_{V_1} z~, \nonumber \\
 \bar{D}_{A_1V_2}(z) &=&  g_{A_2}^* \kappa_A n_{V_2} z^2/2 .
\end{eqnarray}
The other coefficients, which are not obtained
from the symmetry, are zero. The mean numbers of initial chaotic phonons
are denoted as $ n_{V_1} $ and $ n_{V_2} $ for modes $ V_1 $
and $ V_2 $, respectively.

Variances of the integrated intensity for both single and
compound modes can be expressed using relations (17), (18) and
(25):
\begin{eqnarray}            
 \langle (\Delta W_{S_1}(z))^2 \rangle &=& 2 |g_{S_1}|^2
 (n_{V_1} + 1 ) |\xi_{S_1}|^2 z^2 , \nonumber \\
 \langle (\Delta W_{A_1}(z))^2 \rangle &=& 2 |g_{A_1}|^2
 n_{V_1} |\xi_{A_1}|^2 z^2   , \nonumber \\
 \langle (\Delta W_{V_1}(z))^2 \rangle &=& n_{V_1}^2
 + 2 |g_{S_1}|^2 n_{V_1} (|\xi_{S_1}|^2 + n_{V_1} + 1 ) z^2
 \nonumber \\
 & & \mbox{} + 2 |g_{A_1}|^2 n_{V_1} (|\xi_{A_1}|^2 - n_{V_1}) z^2
 + (2g_{A_1}^*g_{S_1}^* \xi_{S_1}\xi_{A_1} n_{V_1} + \mbox{c.c.})
  , \nonumber \\
 \langle (\Delta W_{S_1A_1}(z))^2 \rangle &=&  2|g_{S_1}|^2
 (n_{V_1} + 1 ) |\xi_{S_1}|^2  z^2 + 2|g_{A_1}|^2
 n_{V_1} |\xi_{A_1}|^2  z^2 \nonumber \\
 & & \mbox{} - [ g_{A_1}g_{S_1}
 (2n_{V_1} + 1 ) \xi_{S_1}^* \xi_{A_1}^*
  z^2 + \mbox{c.c.} ] , \nonumber \\
 \langle (\Delta W_{S_1V_1}(z))^2 \rangle &=& n_{V_1}^2 +
 2 |g_{S_1}|^2 ( 3|\xi_{S_1}|^2+ n_{V_1} + 1) (n_{V_1} + 1 ) z^2
 \nonumber \\
 & & \mbox{} +  [ 2g_{A_1}g_{S_1}
 (n_{V_1} + 1 ) \xi_{S_1}^* \xi_{A_1}^*
  z^2 + \mbox{c.c.} ], \nonumber \\
 \langle (\Delta W_{A_1V_1}(z))^2 \rangle &=& n_{V_1}^2 +
  2 |g_{A_1}|^2 n_{V_1} (n_{V_1} - |\xi_{A_1}|^2) z^2 \nonumber \\
  & & \mbox{} -  [ 2g_{A_1}g_{S_1}
  n_{V_1} \xi_{S_1}^* \xi_{A_1}^*
  z^2 + \mbox{c.c.} ].
\end{eqnarray}
The above short-length expressions (26) indicate negative variances of the
integrated intensity only in
the compound mode ($ S_1,A_1 $).
The influence of linear Stokes ($ \kappa_S
$) and anti-Stokes ($ \kappa_A $) couplings on the variances cannot
be deduced from the expressions restricted up to $ z^2 $.

As has been discussed previously, squeezing of vacuum
fluctuations cannot be obtained in single modes.
However, the short-length solution provides interesting
results for principal squeeze variances of the following
compound modes, using (20) and (25):
\begin{eqnarray}             
 \lambda_{S_1A_1}(z) &=& 2[ 1 + |g_{S_1}|^2(n_{V_1}+1) z^2 +
  |g_{A_1}|^2 n_{V_1} z^2 - |g_{A_1}||g_{S_1}| (1+ 2n_{V_1}) z^2] ,
  \nonumber \\
 \lambda_{S_1V_1}(z) &=& 2[ 1 + n_{V_1} - 2|g_{S_1}|(n_{V_1}+1)
z~+ 2|g_{S_1}|^2(n_{V_1}+1) z^2 - |g_{A_1}|^2 n_{V_1} z^2] ,
  \nonumber \\
 \lambda_{A_1V_1}(z) &=& 2[ 1 + n_{V_1} - (ig_{A_1}^* n_{V_1}
z~+ \mbox{c.c.} ) + |g_{S_1}|^2(n_{V_1}+1) z^2 ],  \nonumber \\
 \lambda_{S_1V_2}(z) &=& 2[ 1 + n_{V_2} + |g_{S_1}|^2(n_{V_1}+1) z^2 +
  |g_{S_2}|^2(n_{V_2}+1) z^2 - |g_{A_2}|^2 n_{V_2} z^2 \nonumber \\
  & & \mbox{} - |g_{S_2}|\kappa_S|(n_{V_2}+1) z^2] ,
  \nonumber \\
 \lambda_{A_1V_2}(z) &=& 2[ 1 + n_{V_2} + |g_{S_2}|^2(n_{V_2}+1) z^2 +
  |g_{A_1}|^2 n_{V_1} z^2 - |g_{A_2}|^2 n_{V_2} z^2 \nonumber \\
  & & \mbox{} -
  (g_{A_2}^*\kappa_A n_{V_2} z^2 + \mbox{c.c.}) ] .
\end{eqnarray}
Thus, squeezing can be reached in mode ($ S_1,A_1 $) for
small $ n_{V_1} $.
Especially, the above expresions show that Stokes linear coupling
($ \kappa_S $) could generate
squeezing in mode ($ S_1,V_2 $) and anti-Stokes linear coupling
($ \kappa_A $) could lead to squeezing in mode ($ A_1,V_2 $).

The advantage of the short-length solutions stemms from their
simpler expressions which can be relatively easily analyzed.
The relations above introduced enable us to reveal partly
the role of Stokes and anti-Stokes coupling constants in
generation of light with nonclassical properties. As it will
be seen in the next section, the dynamics of the statistical properties
is more complex in general than it may be seen in the
short-length solution; but all tendencies are obvious from the
short-length solution. The short-length solution is important
especially for understanding the role of various initial phases,
which can substantially change the statistical behaviour of nonlinear
couplers.

\section{Statistical behaviour of nonlinear couplers}

This section contains a discussion of the statistical
behaviour of the Raman and Brillouin couplers obtained on the basis of
numerical resuls gained by applying the method described
in section 2.1. In special cases, analytical solution
from section 2.2 can be used to verify the correctness
of the numerical approach. The discussion further generalizes results
obtained in the previous section by means of the short-length
solution. The couplers with Brillouin and Raman processes in
operation will be discussed separately.

Photon number distribution $ p(n_j,z) $, reduced
moments of the integrated intensity $ \langle W^k (z)\rangle
/ $ $ \langle W(z) \rangle^k - 1 $, variances of quadrature
components $ \langle (\Delta \hat{p}_j(z))^2 \rangle $,
$ \langle (\Delta \hat{q}_j(z))^2 \rangle $, principal
squeeze variance $ \lambda_j(z) $, and the uncertainty product
parameter $ u_j(z) $ (for definitions, see \cite{pp4})
will be used when discussing the statistical properties of
single and compound modes. Nonclassical behaviour of
a field exhibits itself by sub-Poissonian photon number
statistics, by negative reduced moments
of the integrated intensity (they indicate antibunching
of photons; a field with negative second reduced
moment of the integrated intensity possesses a sub-Poissonian
statistics) or by squeezing of quadrature variances (the variance is squeezed
if its value is under 1 (2) for a single (compound) mode).

Squeezing of vacuum fluctuations and sub-Poissonian statistics
cannot be reached in single modes, if the fields are initially
in coherent states or in coherent states with superimposed
noise (this is connected with the fact that the coefficients
$ C_j(z) $ in the normal characteristic function (14)
are zero provided that they are initially zero). That is why
we orientate our attention mainly to two-mode fields.

We choose the condition $ |g_A| > |g_S| $ in cases discussed
below, because such a condition is suitable for nonclassical
light generation \cite{pie1,pie2}.

\subsection{Coupler with Brillouin processes}

We suppose initial coherent states in all six modes of
the coupler. We divide the discussion to six sections;
in each section a special configuration is discussed.
These configurations change from simpler to more complex
ones.

\vspace{3mm}
\noindent
{\it Influence of $ \kappa_S $ on stimulated Stokes process}
\vspace{2mm}

We suppose stimulated Stokes and spontaneous
anti-Stokes processes in waveguide 1 ($ g_{S_1} \neq 0 $,
$ g_{A_1} \neq 0 $, $ \xi_{S_1} \neq 0 $, $ \xi_{V_1} \neq 0 $,
$ \xi_{A_1} = 0 $).
Reduced moments of the integrated intensity (in the following
we simply speak of moments of intensity) for compound mode
($ S_1,V_1 $) can be negative if phases are suitably
chosen (see the short-length solution (23)). Moments
of intensity in modes  ($ S_1,A_1 $) and ($ A_1,V_1 $)
are non-negative. Squeezing
can be reached in modes ($ S_1,A_1 $) and ($ S_1,V_1 $). There is
a periodicity in the spatial development with the period
$ 1/\sqrt{|g_{A_1}|^2 - |g_{S_1}|^2 } $ (see \cite{pie1}).
Single and compound modes return periodically to coherent
states (see Fig.~2 for mode ($ S_1,A_1 $)).
\begin{figure}        
  \centerline{\hbox{\psfig{file=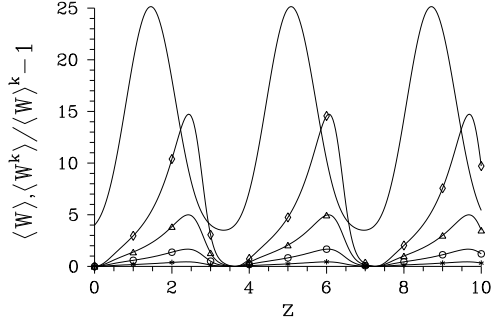,height=4.2cm}
     }}
  \vspace{3mm}
  \caption{Mean integrated intensity $ \langle W(z)\rangle $
 (solid curve without
 denotation) and reduced moments of the integrated intensity
 $ \langle W^k(z)\rangle/\langle
 W(z)\rangle^k -1 $ for $ k=2 (\ast) $, $ k=3 (\circ) $, $ k=4 (\triangle)$,
 and $ k=5 (\diamond) $ for mode ($ S_1,A_1 $);
 $ g_{S_1} = 1 $, $ g_{A_1} = 2 $, $ \xi_{S_1} = 2i $,
 $ \xi_{V_1} = 1 $, and the other parameters are zero.}
\end{figure}

The introduction of the linear Stokes coupling ($ \kappa_S \neq 0 $,
$ \xi_{S_2} \neq 0 $) leads to the occurrence of negative
moments of intensity also in modes ($ S_1,A_1 $)
(see Fig. 3a, compare it with Fig. 2), ($ S_2,V_1 $),
and ($ S_2,A_1 $). Fig. 3b illustrates a typical spatial
development of the photon number distribution of compound
modes with negative moments of intensity: the initially
Poissonian statistics develope to regions exhibiting sub-Poissonian
statistics in interchange with regions exhibiting super-Poissonian
statistics.
Squeezing of vacuum fluctuations can be reached also in modes
($ S_2,V_1 $) and ($ S_2,A_1 $) (see Fig. 3c for mode ($ S_2,A_1 $)).
The increase of $ \kappa_S $ decreases the
period of the spatial dynamics (as indicated by the analytical
solution in (11)) and it does not lead to an increase
of noise for longer $ z $, but to a tendency to conserve the
initial statistics.
\begin{figure}        
  \centerline{\hbox{(a) \hspace{5mm}\psfig{file=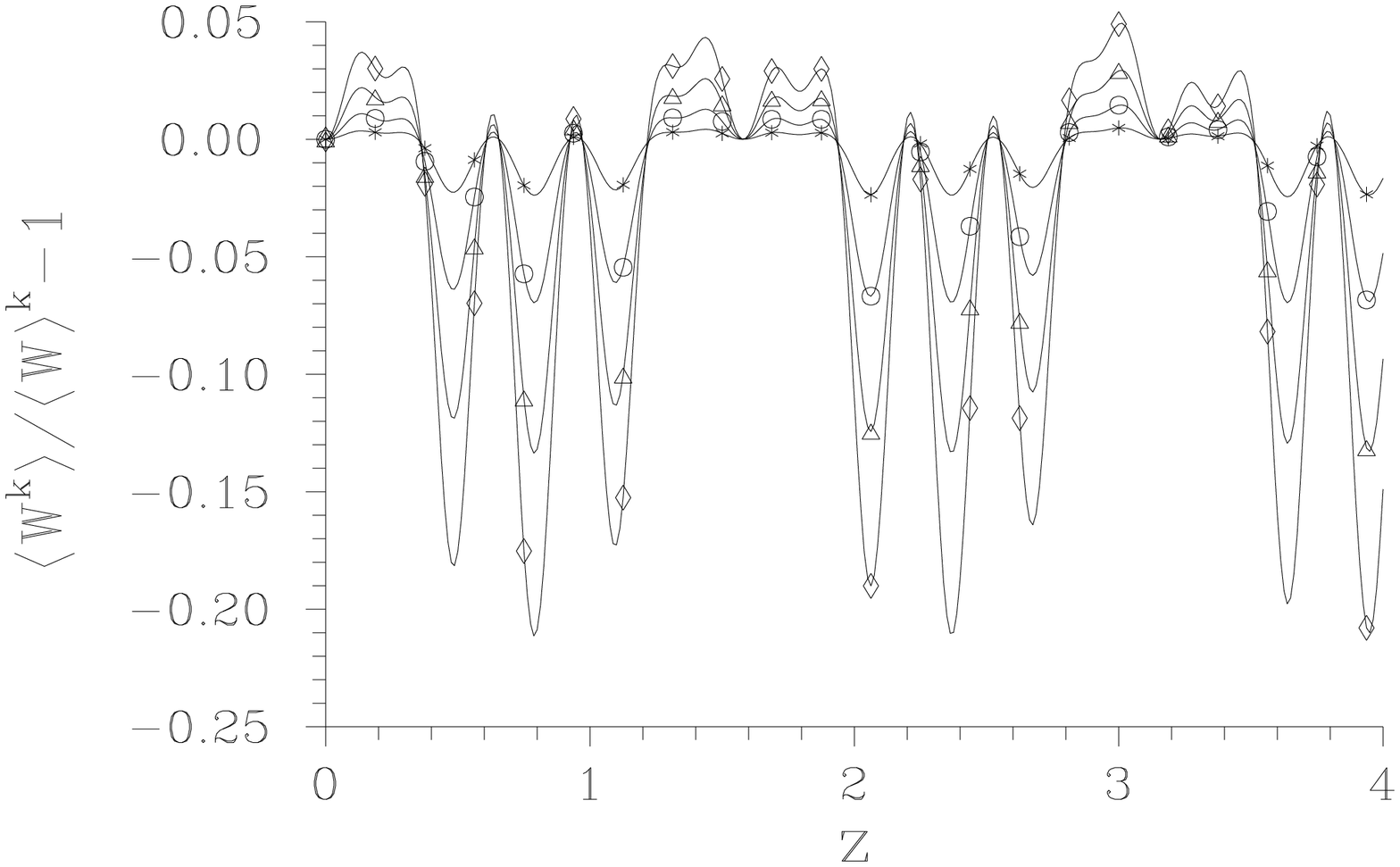,height=4.2cm}
     }}
  \vspace{5mm}
  \centerline{\hbox{(b) \hspace{13mm} \psfig{file=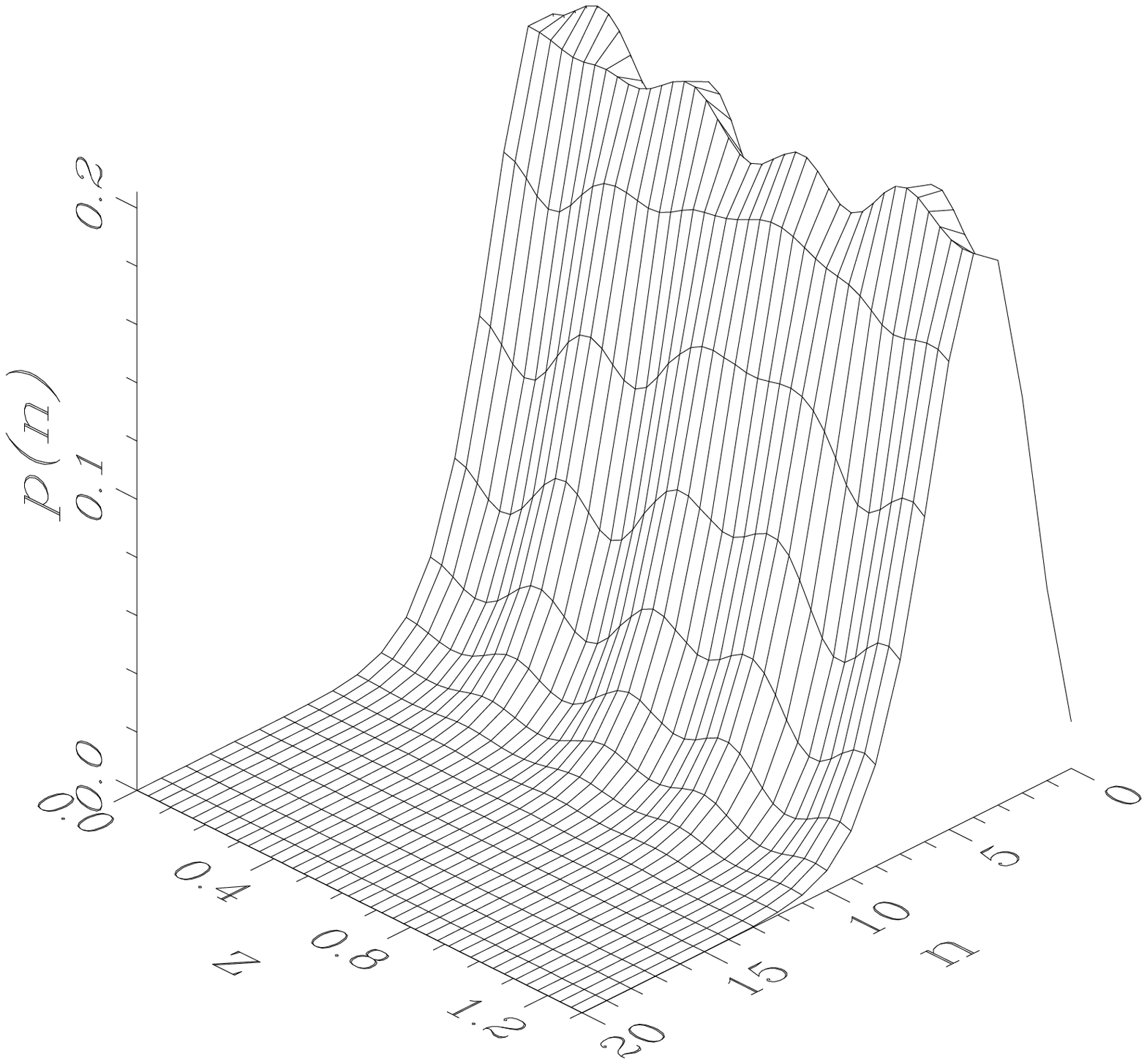,height=4.5cm}
     \hspace{8mm} \mbox{} }}
  \vspace{5mm}
  \centerline{\hbox{(c) \hspace{5mm}\psfig{file=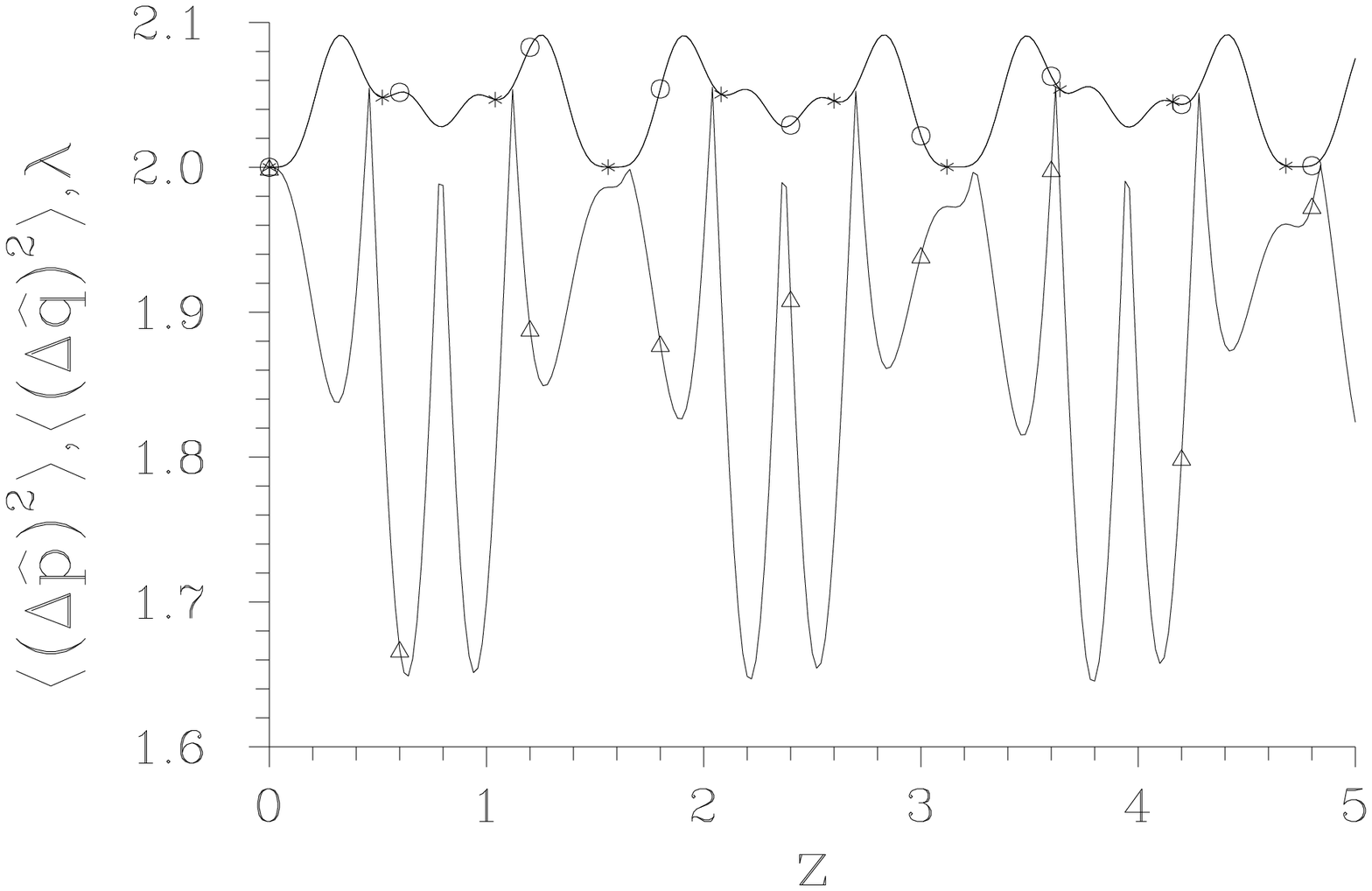,height=4.2cm}
     }}
  \vspace{3mm}
  \caption{Reduced moments of the integrated intensity
 $ \langle W^k(z)\rangle/\langle
 W(z)\rangle^k -1 $ for $ k=2 (\ast) $, $ k=3 (\circ) $, $ k=4 (\triangle)$,
 and $ k=5 (\diamond) $ for mode ($ S_1,A_1 $) {\bf (a)}, photon
 number distribution $ p(n,z) $ for mode ($ S_1,A_1 $) {\bf (b)},
 and quadrature variances $ \langle(\Delta
 \hat{p}(z))^2\rangle (\ast) $, $ \langle (\Delta \hat{q}(z))^2\rangle
 (\circ) $, and principal
 squeeze variance $ \lambda(z) (\triangle) $
 {\bf (c)} for mode ($ S_2,A_1 $);
 $ g_{S_1} = 1 $, $ g_{A_1} = 2 $, $ \kappa_S = -10 $, $ \xi_{S_1} = 2 $,
 $ \xi_{V_1} = 1 $, $ \xi_{S_2} = 2 $, and the other parameters are zero.}
\end{figure}

The introduction of the Stokes process in waveguide 2
($ g_{S_2} \neq 0 $, $ \xi_{V_2} \neq 0 $) does not lead
to nonclassical properties of mode ($ V_1,V_2 $) at all.

\vspace{3mm}
\noindent
{\it Influence of $ \kappa_A $ on stimulated anti-Stokes process}
\vspace{2mm}

We suppose stimulated anti-Stokes and spontaneous
Stokes processes in waveguide 1 ($ g_{S_1} \neq 0 $,
$ g_{A_1} \neq 0 $, $ \xi_{A_1} \neq 0 $, $ \xi_{V_1} \neq 0 $,
$ \xi_{S_1} = 0 $).
Negative moments of intensity can be reached in mode
($ S_1,A_1 $) if the initial phase of $ \xi_{A_1} $
is suitably chosen ($ \arg(\xi_{A_1}) = \pm \pi/2 $).
Squeezing in modes ($ S_1,A_1 $) and ($ S_1,V_1 $)
reflects nonclassical properties of fields in this case.
Non-zero anti-Stokes coupling ($ \kappa_A \neq 0 $, $ \xi_{A_2}
\neq 0 $) does not support nonclassical properties,
but it can lead to transmission of light exhibiting
negative moments of intensity
from mode ($ S_1,A_1 $) to mode ($ S_1,A_2 $).
Non-zero $ \kappa_A $ causes a fast increase of mean
intensity and its moments. Faster oscillations in $ z $ occur in
the spatial development of discussed quantities, but they are
connected with a gradual increase of noise.
There occur periods with
a noise reduction.
The coupling constant $ \kappa_A $ destroys gradually
squeezing in modes ($ S_1,A_1 $) and ($ S_1,V_1 $).

\vspace{3mm}
\noindent
{\it Influence of $ \kappa_S $ on stimulated Stokes and
anti-Stokes processes}
\vspace{2mm}

Stimulated Stokes and anti-Stokes processes in waveguide 1 are
assumed
($ g_{S_1} \neq 0 $, $ g_{A_1} \neq 0 $, $ \xi_{S_1} \neq 0 $,
$ \xi_{A_1} \neq 0 $, $ \xi_{V_1} \neq 0 $). Negative moments of
intensity and squeezing occur in both modes ($ S_1,A_1 $) and
($ S_1,V_1 $) if phases are suitably chosen ($ \phi_V + \phi_S
- \phi_{g_S} = -\pi/2 $, $ \phi_V - \phi_A
+ \phi_{g_A} = -\pi/2 $, see \cite{pie1}).

The interaction of mode $ S_2 $ through Stokes coupling
($ \kappa_S \neq 0 $, $ \xi_{S_2} \neq 0 $) leads to
the occurrence of negative moments of the intensity
and to squeezing also in modes ($ S_2,A_1 $) (see Figs. 4a
and 4b) and
($ S_2,V_1 $). Fig. 4b demonstrates photon number distribution
$ p(n,z) $ for mode ($ S_2,A_1 $), which exhibits oscillations between
sub-Poissonian and super-Poissonian statistics when $ z $
increases (compare Fig. 4b with Fig. 4a).
\begin{figure}        
  \centerline{\hbox{(a) \hspace{5mm}\psfig{file=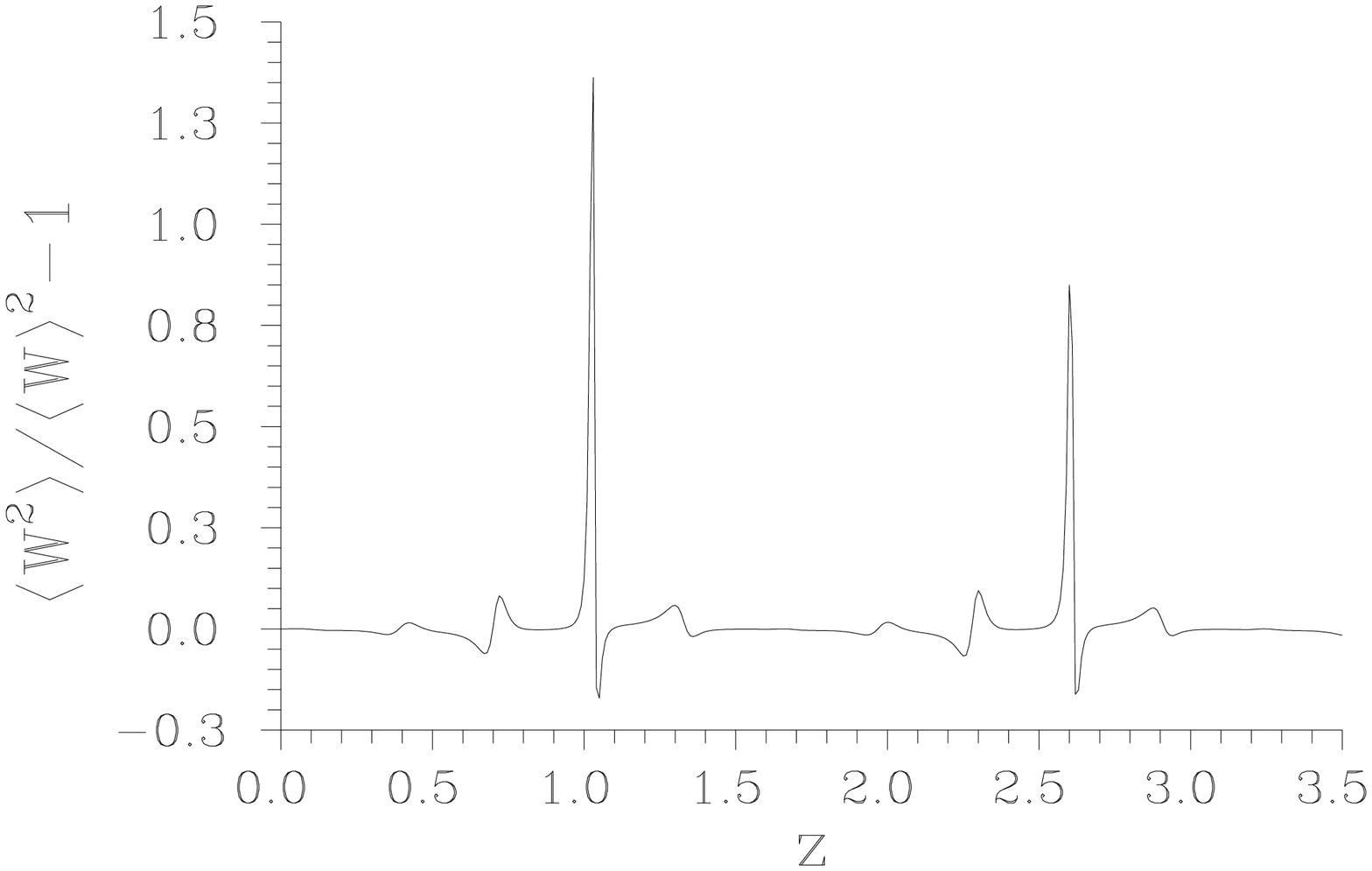,height=4.2cm}
     }}
  \vspace{5mm}
  \centerline{\hbox{(b) \hspace{13mm} \psfig{file=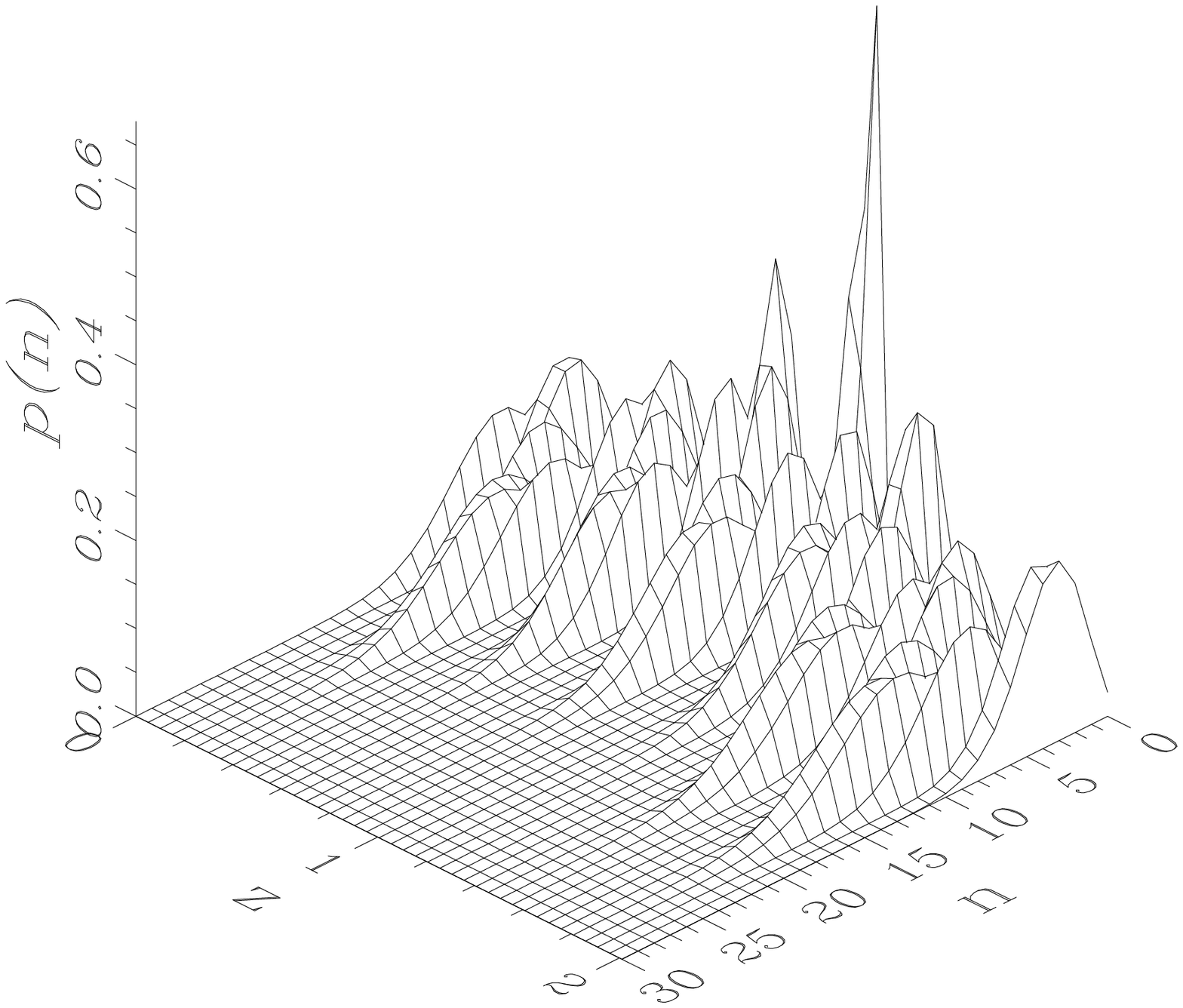,height=4.5cm}
     \hspace{8mm} \mbox{} }}
  \vspace{3mm}
  \caption{The second reduced moment of the integrated intensity
 $ \langle W^2(z)\rangle/\langle
 W(z)\rangle^2 -1 $ {\bf (a)}
 and the photon number distribution $ p(n,z) $ for mode
 ($ S_2,A_1 $) {\bf (b)};
 $ g_{S_1} = 1 $, $ g_{A_1} = 2 $, $ \kappa_S = -10 $,
 $ \xi_{S_1} = -2i $, $ \xi_{A_1} = 2i $,
 $ \xi_{V_1} = 1 $,  $ \xi_{S_2} = 2 $,
 and the other parameters are zero.}
\end{figure}

If the phases are not suitably chosen, moments of mode
($ S_1,A_1 $) are non-negative when the Stokes coupling
is not included ($ \kappa_S = 0 $) ($ \phi_V + \phi_S
- \phi_{g_S} = -\pi/2 $, $ \phi_V - \phi_A
+ \phi_{g_A} = \pi/2 $, see \cite{pie1}).
The inclusion of the Stokes coupling can stimulate negative
moments of intensity in this mode.

The addition of stimulated Stokes process in waveguide 2
($ g_{S_2} \neq 0 $, $ \xi_{S_2} \neq 0 $) does not
lead to negative moments of intensity in mode
($ V_1,V_2 $).

\newpage

\vspace{3mm}
\noindent
{\it Influence of $ \kappa_A $ on stimulated Stokes and
anti-Stokes processes}
\vspace{2mm}

Non-zero anti-Stokes coupling ($ \kappa_A \neq 0 $, $ \xi_{A_2}
\neq 0 $) of waveguide 2 suppresses nonclassical behaviour
of compound modes in waveguide 1
($ g_{S_1} \neq 0 $, $ g_{A_1} \neq 0 $, $ \xi_{S_1} \neq 0 $,
$ \xi_{A_1} \neq 0 $, $ \xi_{V_1} \neq 0 $) for longer $ z $.
It introduces oscillations to the spatial development of
quantities under investigation
and it increases noise in moments of intensity
and supports higher values of quadrature variances. However,
negative moments of intensity and squeezing for shorter $ z $
can be reached in mode ($ S_1,A_2 $) as a consequence of the
linear anti-Stokes coupling.

The anti-Stokes process in waveguide 2
($ g_{A_2} \neq 0 $, $ \xi_{V_2} \neq 0 $)
does not create nonclassical effects in compound modes
($ A_1,A_2 $), ($ A_2,V_1 $), ($ A_1,V_2 $), ($ V_1,V_2 $),
and ($ S_1,V_2 $).

\vspace{3mm}
\noindent
{\it Influence of $ \kappa_S $ on stimulated Stokes and
anti-Stokes processes in both waveguides}
\vspace{2mm}

Stimulated Stokes and anti-Stokes processes in both the waveguides
are assumed now ($ g_{S_1} \neq 0 $, $ g_{A_1} \neq 0 $, $ \xi_{S_1} \neq 0 $,
$ \xi_{A_1} \neq 0 $, $ \xi_{V_1} \neq 0 $,
$ g_{S_2} \neq 0 $, $ g_{A_2} \neq 0 $, $ \xi_{S_2} \neq 0 $,
$ \xi_{A_2} \neq 0 $, $ \xi_{V_2} \neq 0 $).
Negative moments of intensity and squeezing occur in
compound modes ($ S_1,A_1 $), ($ S_1,V_1 $), ($ S_2,A_2 $),
and ($ S_2,V_2 $)  when
phases are suitably chosen.
The linear Stokes coupling ($ \kappa_S \neq 0 $) conserves
slightly negative moments of intensity and squeezing in these modes
(see Fig. 5a for mode ($ S_1,A_1 $)) and it can induce
negative moments of intensity and squeezing also in
compound modes composed of single modes in different
waveguides, especially in modes ($ S_2,A_1 $), ($ S_2,V_1 $),
($ S_1,A_2 $), and ($ S_1,V_2 $) (see Fig. 5b for mode
($ S_2,V_1 $); in mode ($ S_1,A_2 $) substantial squeezing of
vacuum fluctuations can be obtained).
The dynamics of the mean intensity and of moments in
Fig. 5a shows a typical behaviour of compound modes;
regions with slightly negative moments are
followed by short regions of high increase of noise,
when a decrease of the mean intensity changes to an
increase. Complex values of $ \kappa_S $ introduce
asymmetry between the waveguides, which affects strongly the
spatial development of the compound modes.
\begin{figure}        
  \centerline{\hbox{(a) \hspace{9mm}\psfig{file=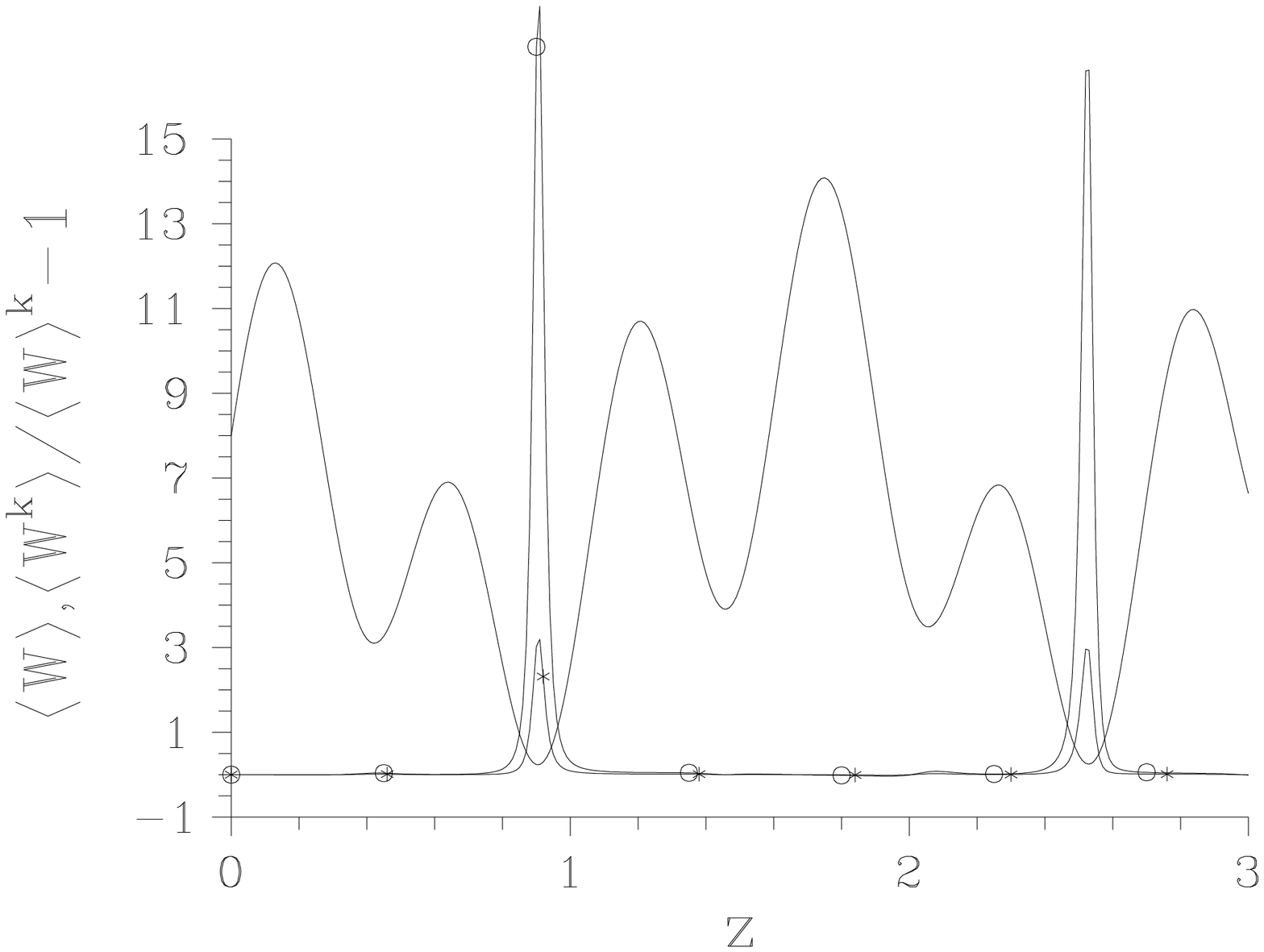,height=4.2cm}
     \hspace{4mm} \mbox{} }}
  \vspace{5mm}
  \centerline{\hbox{(b) \hspace{5mm} \psfig{file=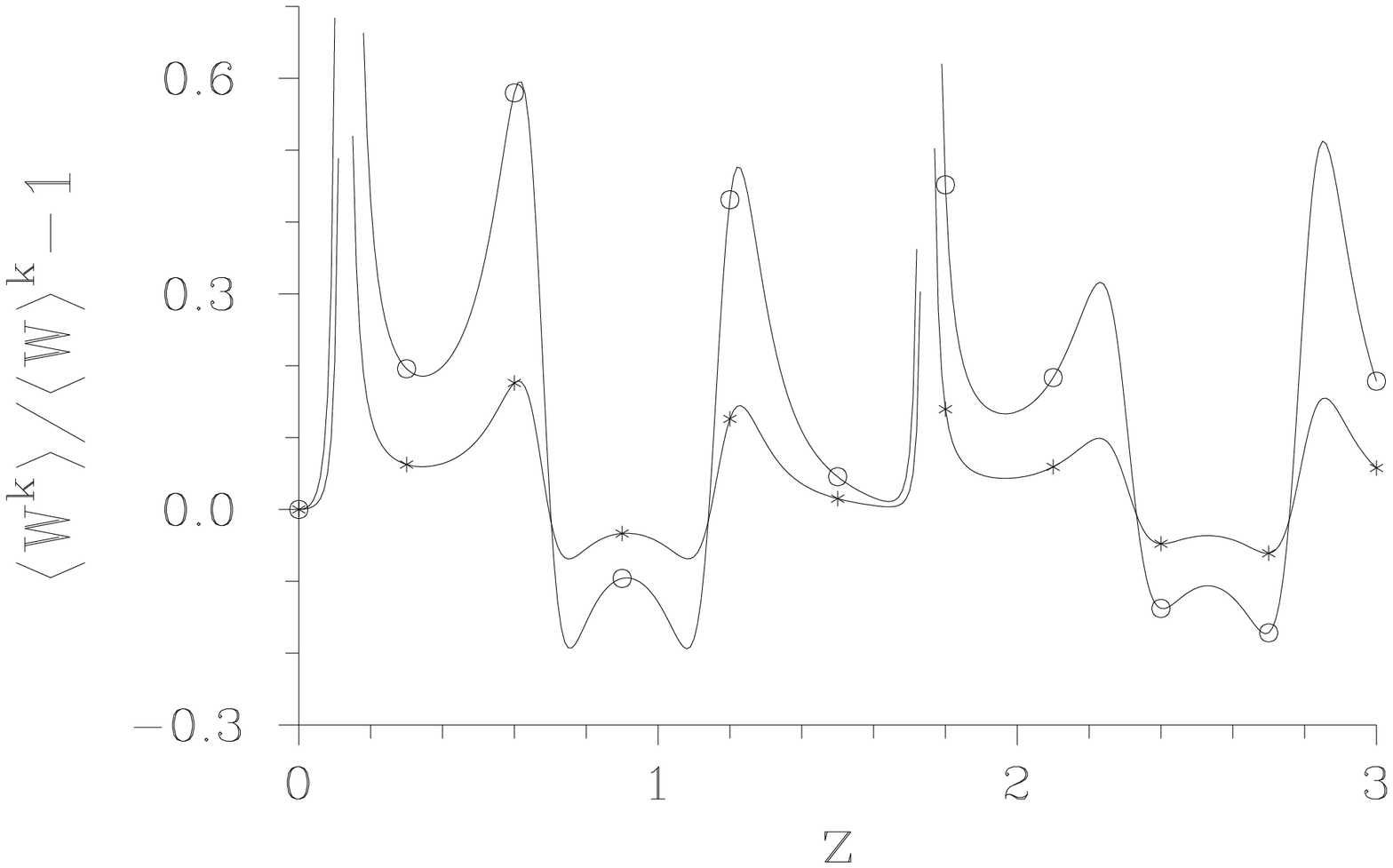,height=4.2cm}
     }}
  \vspace{5mm}
  \centerline{\hbox{(c) \hspace{5mm} \psfig{file=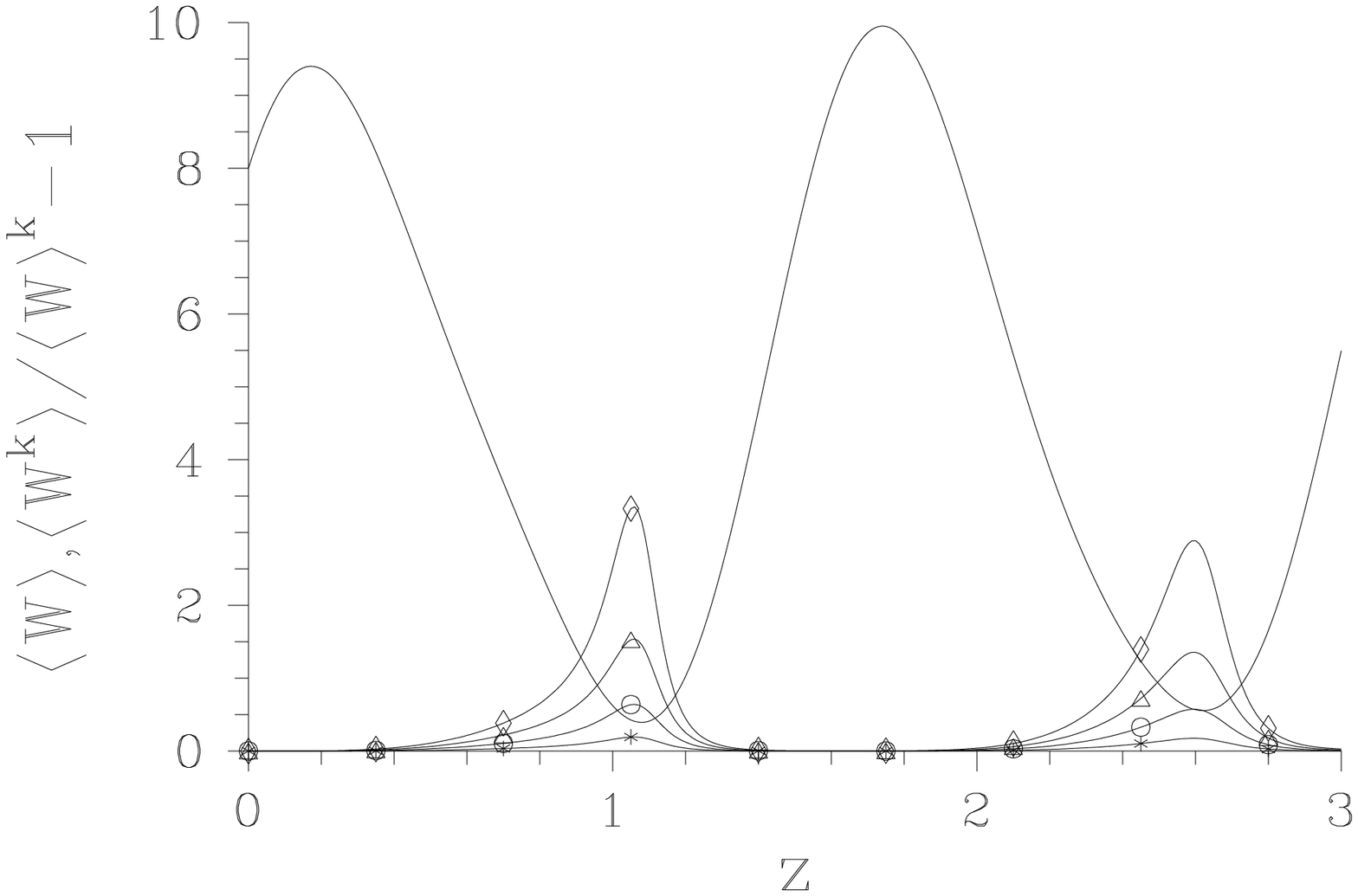,height=4.2cm}
     }}
  \vspace{3mm}
  \caption{Mean integrated intensity $ \langle W(z)\rangle $
 (solid curve without
 denotation) and reduced moments of the integrated intensity
 $ \langle W^k(z)\rangle/\langle
 W(z)\rangle^k -1 $ for $ k=2 (\ast) $ and $ k=3 (\circ) $ for
 mode ($ S_1,A_1 $) {\bf (a)},
 reduced moments of the integrated intensity
 $ \langle W^k(z)\rangle/\langle
 W(z)\rangle^k -1 $ for $ k=2 (\ast) $ and $ k=3 (\circ) $
 for mode ($ S_2,V_1 $) {\bf (b)},
 and
 mean integrated intensity $ \langle W(z)\rangle $
 (solid curve without
 denotation) and reduced moments of the integrated intensity
 $ \langle W^k(z)\rangle/\langle
 W(z)\rangle^k -1 $ for $ k=2 (\ast) $, $ k=3 (\circ) $, $ k=4 (\triangle)$,
 and $ k=5 (\diamond) $ for mode ($ A_1,A_2 $) {\bf (c)};
 $ g_{S_1} = 1 $, $ g_{A_1} = 2 $, $ \kappa_S = 6i $,
 $ g_{S_2} = 1 $, $ g_{A_2} = 2 $, $ \xi_{S_1} = -2i $,
 $ \xi_{A_1} = 2i $,
 $ \xi_{V_1} = 1 $, $ \xi_{S_2} = -2i $, $ \xi_{A_2} = 2i $,
 $ \xi_{V_2} = 1 $,
 and the other parameters are zero.}
\end{figure}

Negative moments of intensity or squeezing
are not generated in compound modes constituted by the same
single modes in different waveguides, i.e. in modes
($ S_1,S_2 $), ($ A_1,A_2 $), and ($ V_1,V_2 $)
(see Fig. 5c for mode ($ A_1,A_2 $)).
However, these modes possess strong
tendency to return to the coherent states.

\vspace{3mm}
\noindent
{\it Influence of Stokes and anti-Stokes couplings on
 Brillouin processes in both waveguides}
\vspace{2mm}

Now we discuss a general configuration, i.e. we extend
the configuration from the previous part by including
anti-Stokes coupling ($ \kappa_A \neq 0 $). Nonclassical
effects in compound modes occur for the same modes, as discussed
in the previous section, when values of the coupling constant $ \kappa_A $
were small. The increase of values of $ \kappa_A $ causes
a successive increase of moments of intensity for both modes
with negative moments (see Figs. 6a and 6b for
mode ($ S_2,A_1 $)) and modes with non-negative moments.
This means that the
occurrence of negative moments of intensity is restricted
to shorter $ z $ for greater $ \kappa_A $ (compare Figs.
6a and 6b) or negative moments cannot occur at all.

Greater values of $ \kappa_A $ mostly lead to a successive
increase of values of variances
and uncertainty product, but they can also serve to generate
light, squeezing of which gradually develops with $ z $
(see Fig. 6c for mode ($ S_1,A_1 $)).
\begin{figure}        
  \centerline{\hbox{(a) \hspace{5mm} \psfig{file=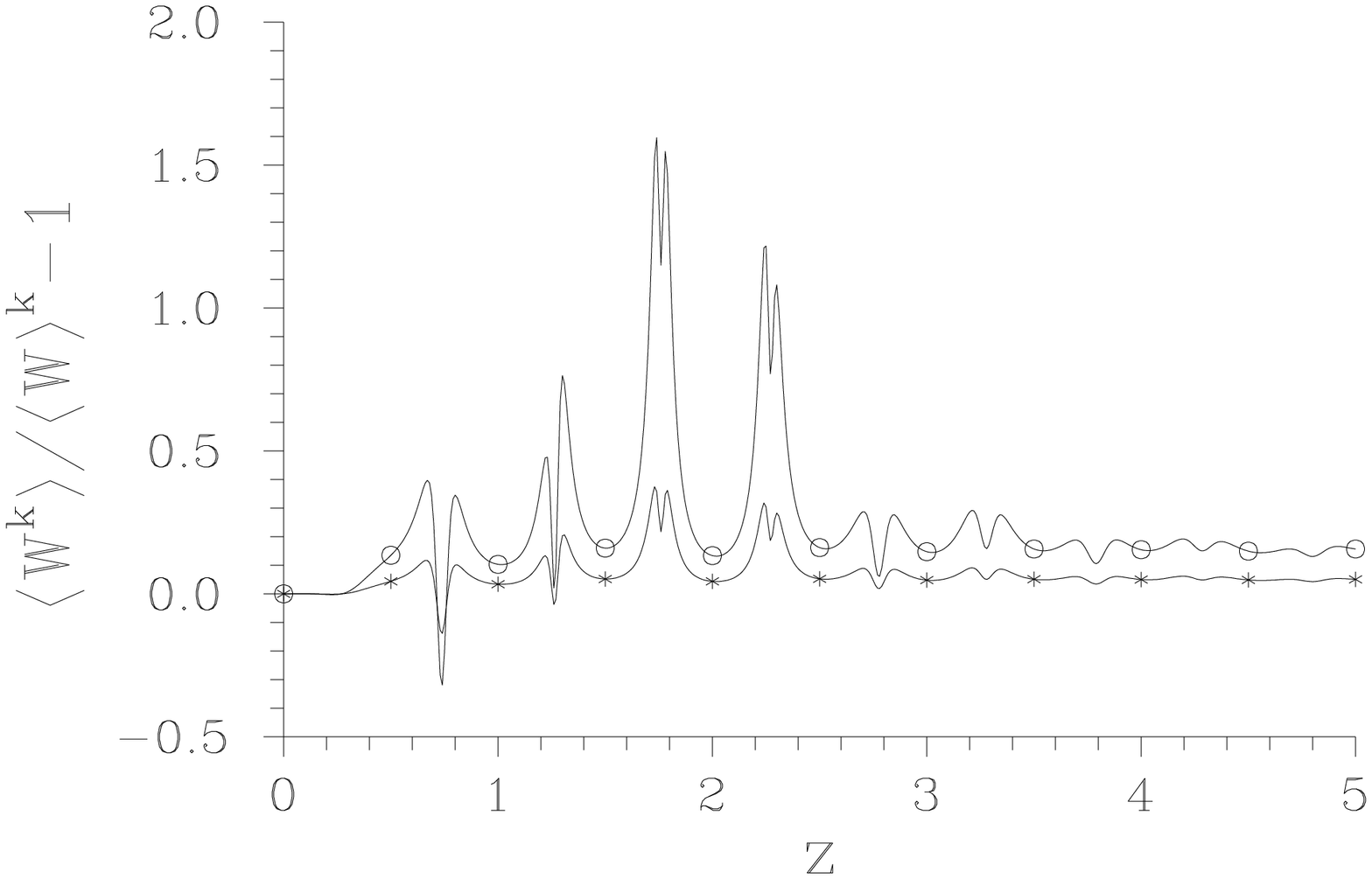,height=4.2cm}
     }}
  \vspace{5mm}
  \centerline{\hbox{(b) \hspace{5mm} \psfig{file=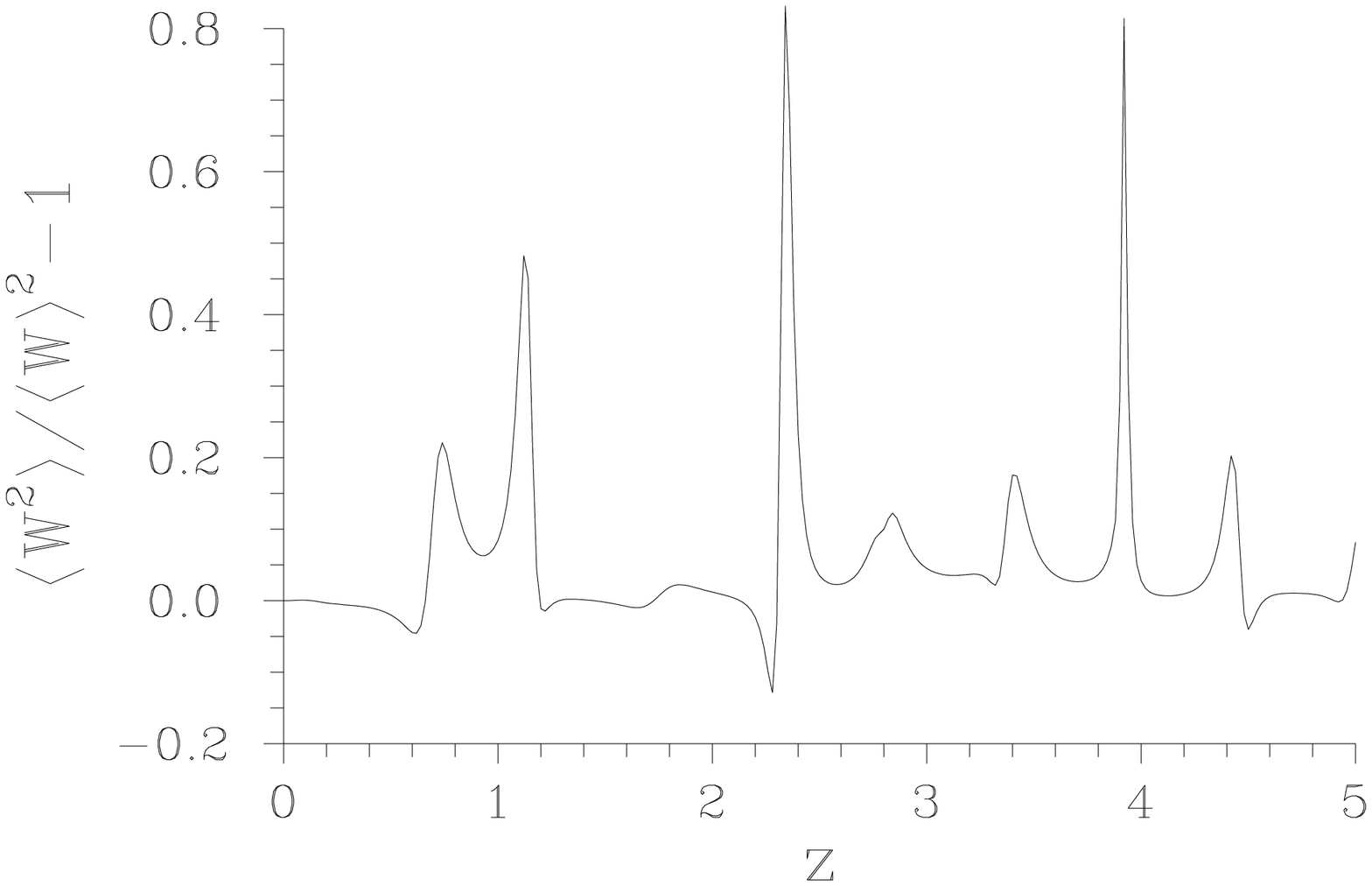,height=4.2cm}
     }}
  \vspace{5mm}
  \centerline{\hbox{(c) \hspace{5mm} \psfig{file=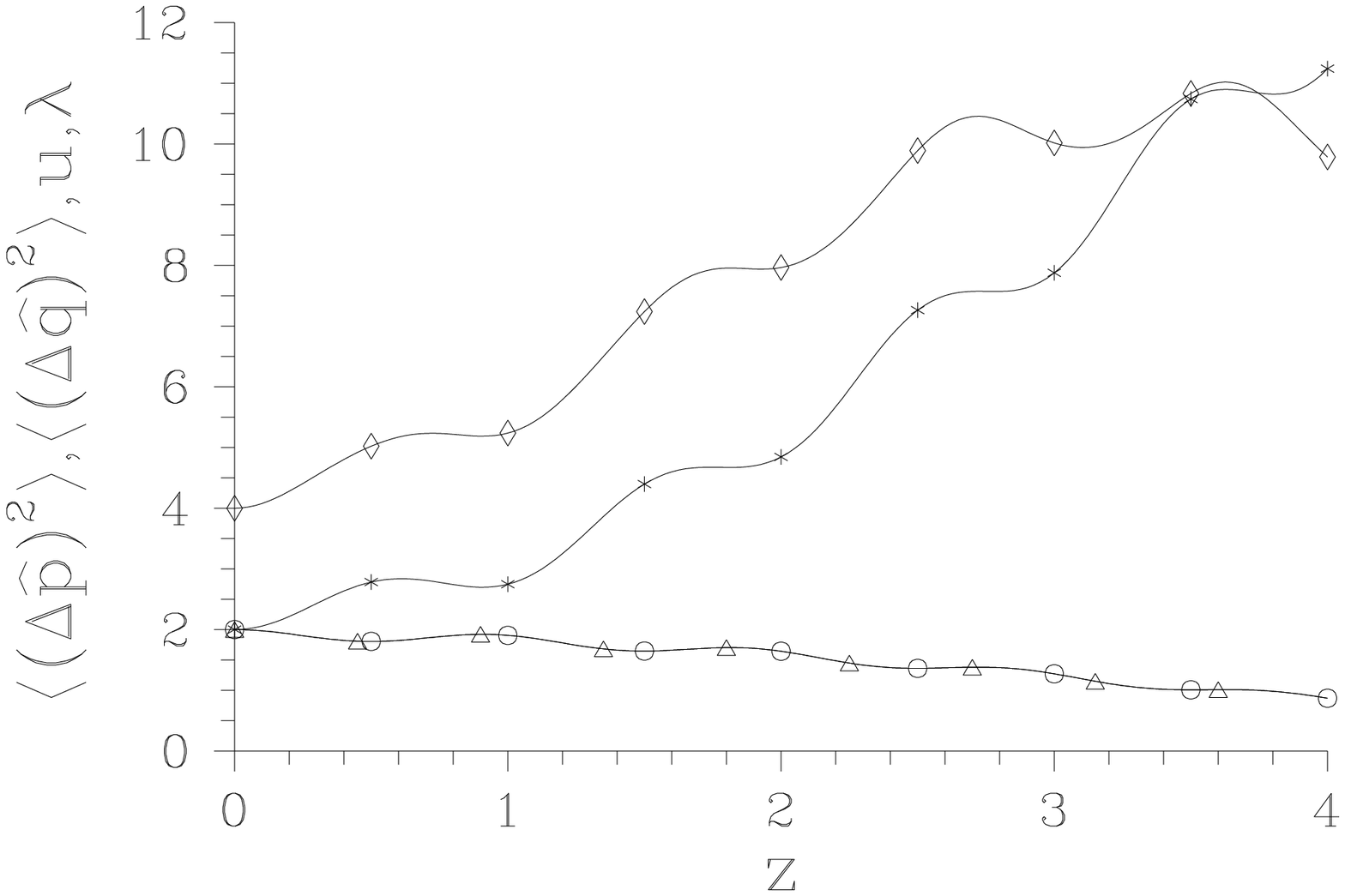,height=4.2cm}
     }}
  \vspace{3mm}
  \caption{Reduced moments of the integrated intensity
 $ \langle W^k(z)\rangle/\langle
 W(z)\rangle^k -1 $ for $ k=2 (\ast) $ and $ k=3 (\circ) $ for
 mode ($ S_2,A_1 $) {\bf (a)},
 the second reduced moment of the integrated intensity
 $ \langle W^2(z)\rangle/\langle W(z)\rangle^2 -1 $
 for mode ($ S_2,A_1 $) ($ \kappa_A = i $) {\bf (b)},
  and
  quadrature variances $ \langle(\Delta
  \hat{p}(z))^2\rangle (\ast) $, $ \langle (\Delta \hat{q}(z))^2\rangle
  (\circ) $, principal
  squeeze variance $ \lambda(z) (\triangle) $, and uncertainty product
  $ u(z) (\diamond) $ for mode
  ($ S_1,A_1 $) {\bf (c)};
 $ \kappa_A = 6i $
 and the other parameters are the same as in Fig. 5.}
\end{figure}

If we compare the above results with the short-length ones, we
can conclude that the general solution reveals richer
possibilities for nonclassical light generation.
The short-length solution together with the analytical
one \cite{pie1} are extremely important for
a suitable choice of phases, which are crucial for
the generation of nonclassical light, as can be deduced
from the short-length solution in the previous section.

\subsection{Coupler with Raman processes}

Initial coherent states in optical modes (Stokes, anti-Stokes)
and initial chaotic states in phonon modes are assumed in the
following discussion, which is again divided into the same
six sections, as was done in the case of Brillouin processes.
In general, the occurrence of regimes giving nonclassical effects in the
statistical behaviour of light is similar as in the
case of Brillouin processes. However, one can expect
a diminished role of various phase relations for the occurrence
of nonclassical light, owing to chaotic statistics of
phonon modes, compared with the above discussed
case. The short-length solution provides
less information in this case.

\vspace{3mm}
\noindent
{\it Influence of $ \kappa_S $ on stimulated Stokes process}
\vspace{2mm}

Stimulated Stokes process ($ g_{S_1} \neq 0 $,
$ g_{A_1} \neq 0 $, $ \xi_{S_1} \neq 0 $, $ n_{V_1} \neq 0 $)
supports the squeezed light generation in compound modes
($ S_1,A_1 $) and ($ S_1,V_1 $). Non-zero Stokes coupling
($ \kappa_S \neq 0 $, $ \xi_{S_2} \neq 0 $) supports
nonclassical light generation. Especially, it induces
antibunching in modes ($ S_1,A_1 $), ($ S_1,V_1 $),
($ S_2,A_1 $), and ($ S_2,V_1 $) and also squeezing
in modes ($ S_2,A_1 $) and ($ S_2,V_1 $). The increase
of values of $ n_{V_1} $ destroys antibunching
in all modes. Antibunching in modes ($ S_2,A_1 $) and ($ S_2,V_1 $)
disappears for smaller $ n_{V_1} $.

\vspace{3mm}
\noindent
{\it Influence of $ \kappa_A $ on stimulated anti-Stokes process}
\vspace{2mm}

Squeezed light in compound modes ($ S_1,A_1 $) and ($ S_1,V_1 $)
can be generated in stimulated anti-Stokes process
($ g_{S_1} \neq 0 $, $ g_{A_1} \neq 0 $, $ \xi_{A_1} \neq 0 $,
$ n_{V_1} \neq 0 $). If values of mean phonon number $ n_{V_1} $ are small,
antibunching in mode ($ S_1,A_1 $) occurs. In this case,
anti-Stokes coupling ($ \kappa_A \neq 0 $, $ \xi_{A_2} \neq 0 $)
leads to transmission of antibunching to mode ($ S_1,A_2 $).
The anti-Stokes linear coupling supports also squeezing in
mode ($ S_1,A_2 $). However, antibunching and squeezing
can occur only provided that $ n_{V_1} $ is sufficiently small.

\vspace{3mm}
\noindent
{\it Influence of $ \kappa_S $ on stimulated Stokes
and anti-Stokes processes}
\vspace{2mm}

Negative moments of intensity in mode ($ S_1,A_1 $)
and squeezing in modes ($ S_1,A_1 $) and ($ S_1,V_1 $)
characterize nonclassical properties of light for
this case ($ g_{S_1} \neq 0 $, $ g_{A_1} \neq 0 $, $ \xi_{S_1} \neq 0 $,
$ \xi_{A_1} \neq 0 $, $ n_{V_1} \neq 0 $).
Fig. 7a shows the photon number distribution $ p(n,z) $
for mode ($ S_1,A_1 $), which developed from Poissonian statistics
through sub-Poissonian statistics to super-Poissonian statistics
and this evolution repeats with increasing $ z $. The corresponding
moments of intensity, demonstrating photon antibunching at the
corresponding $ z $, are given in Fig. 7b.
Squeezing of vacuum fluctuations in mode ($ S_1,V_1 $) can be seen in Fig. 8.
Increasing
values of $ n_{V_1} $ smooth out negative moments of
intensity. Thus, negative moments in mode ($ S_1,V_1 $)
cannot occur owing to chaotic phonon statistics (in the opposite
to Brillouin scattering)
and moments
can evolve periodically and they can reach
the initial values for longer $ z $.
But negative moments of intensity in mode ($ S_1,V_1 $) can be induced by
linear Stokes coupling ($ \kappa_S \neq 0 $,
$ \xi_{S_2} \neq 0 $). Non-zero Stokes coupling provides
negative moments and squeezing
in compound modes composed of single modes in different
waveguides, especiallly in ($ S_2,A_1 $) and ($ S_2,V_1 $).
No antibunching and squeezing are possible in modes
($ S_1,S_2 $) and ($ V_1,V_2 $) ($ g_{S_2} \neq 0 $,
$ g_{A_2} \neq 0 $, $ n_{V_2} \neq 0 $).
\begin{figure}        
  \centerline{\hbox{(a) \hspace{13mm} \psfig{file=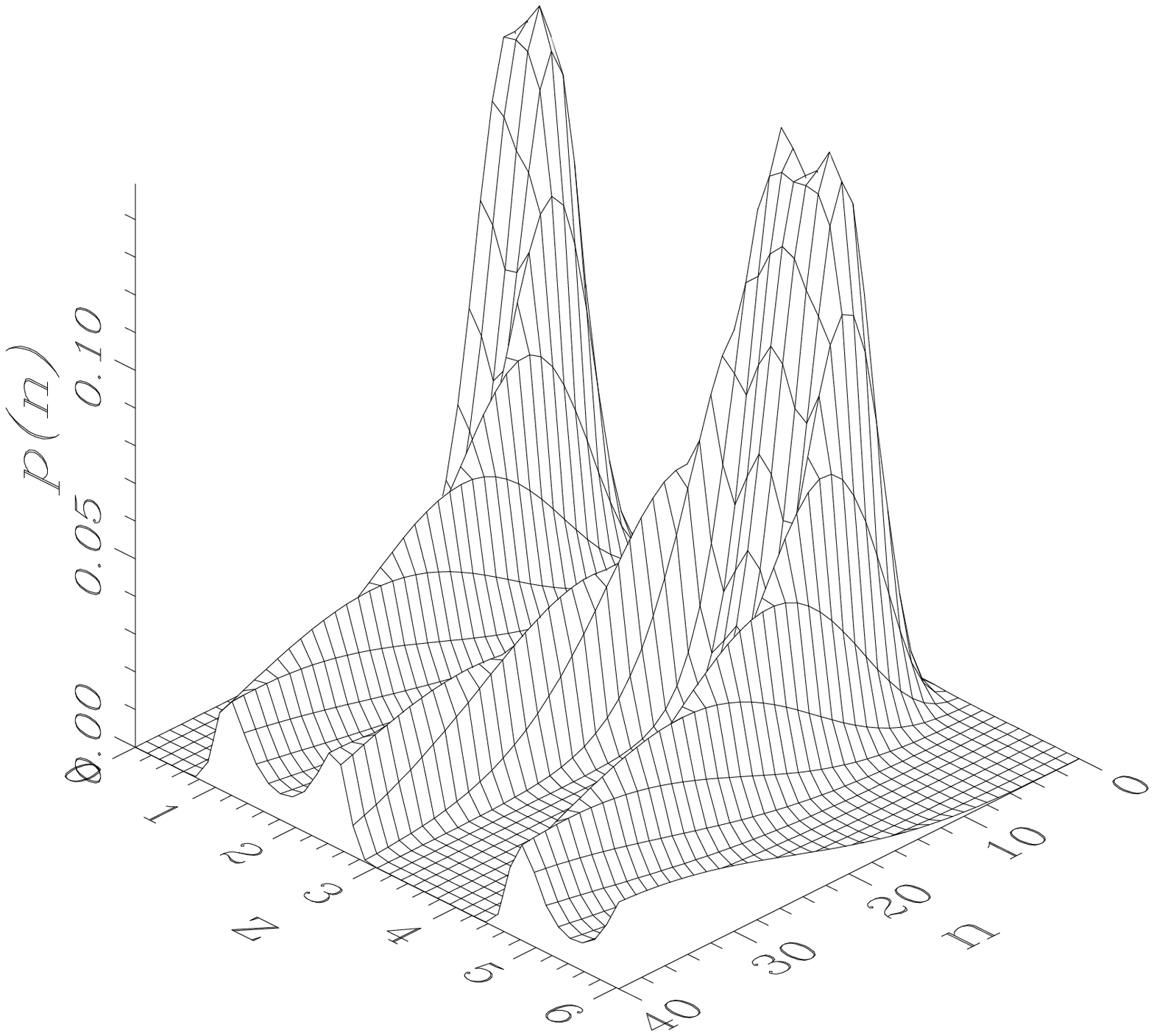,height=4.5cm}
     \hspace{8mm} \mbox{} }}
  \vspace{5mm}
  \centerline{\hbox{(b) \hspace{5mm} \psfig{file=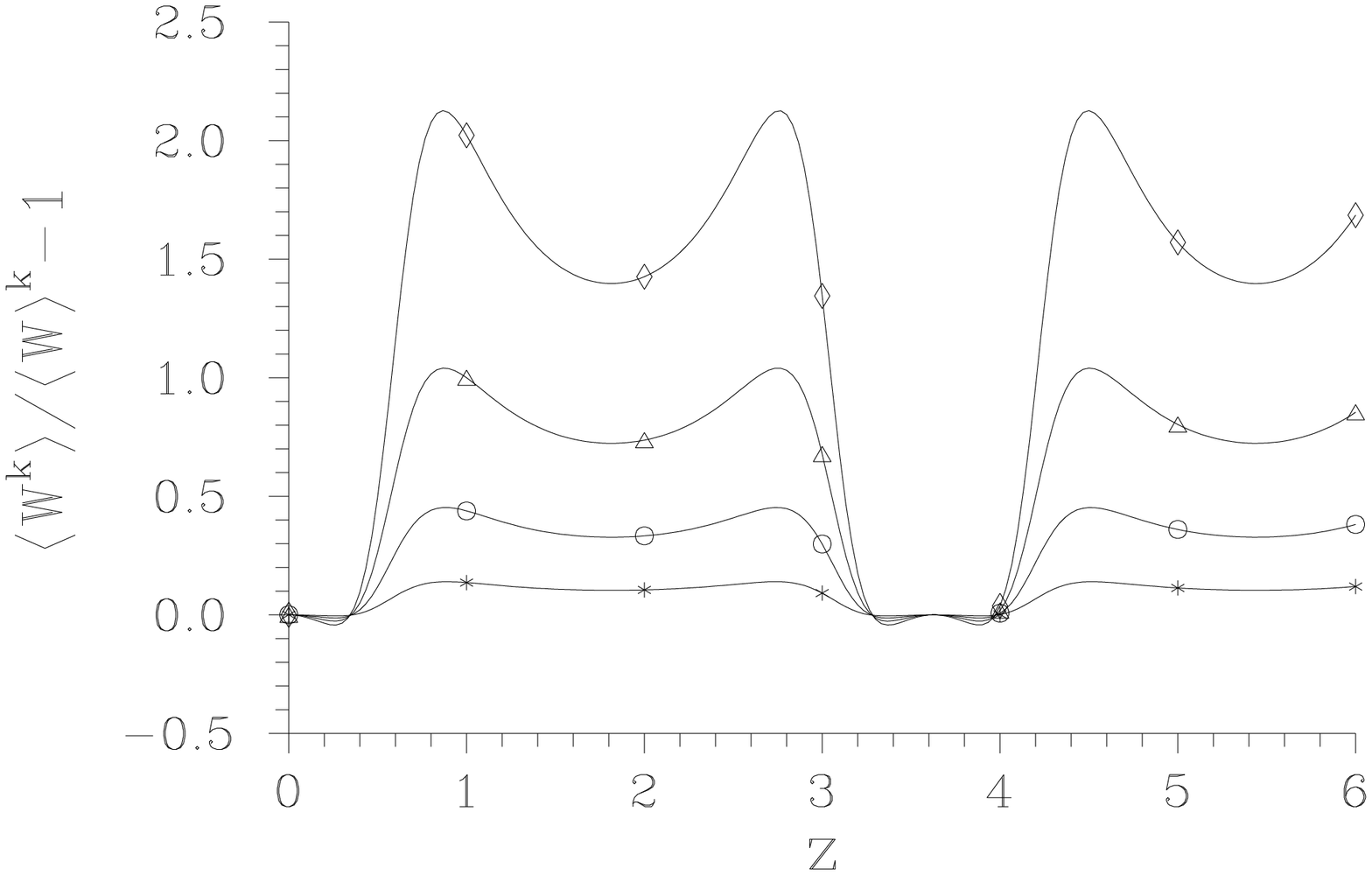,height=4.2cm}
     }}
  \vspace{3mm}
  \caption{Photon number distribution
 $ p(n,z) $ {\bf (a)} and reduced moments of the integrated intensity
 $ \langle W^k(z)\rangle/\langle
 W(z)\rangle^k -1 $ for $ k=2 (\ast) $, $ k=3 (\circ) $, $ k=4 (\triangle)$,
 and $ k=5 (\diamond) $ {\bf (b) } for mode ($ S_1,A_1 $);
 $ g_{S_1} = 1 $, $ g_{A_1} = 2 $, $ \xi_{S_1} = -2i $,
 $ \xi_{A_1} = 2i $,
 $ n_{V_1} = 0.1 $,
 and the other parameters are zero.}
\end{figure}
\begin{figure}        
  \centerline{\hbox{\psfig{file=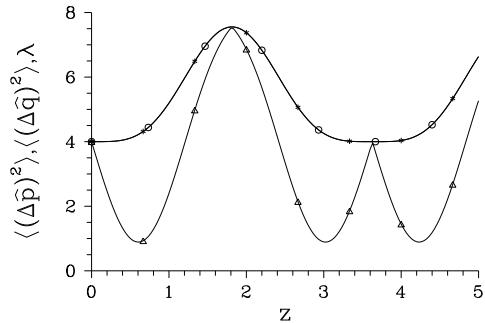,height=4.2cm}
     }}
  \vspace{3mm}
  \caption{Quadrature variances $ \langle(\Delta
 \hat{p}(z))^2\rangle (\ast) $, $ \langle (\Delta \hat{q}(z))^2\rangle
 (\circ) $, and principal squeeze variance $ \lambda(z) (\triangle) $
 for mode ($ S_1,V_1 $);
 $ n_{V_1} = 1 $
 and the other parameters are the same as in Fig. 7.}
\end{figure}

\vspace{3mm}
\noindent
{\it Influence of $ \kappa_A $ on stimulated Stokes and
anti-Stokes processes}
\vspace{2mm}

Similarly as for Brillouin processes,
non-zero anti-Stokes coupling ($ \kappa_A \neq 0 $, $ \xi_{A_2}
\neq 0 $) of waveguide 2 to waveguide 1 with active
stimulated Stokes and anti-Stokes processes
($ g_{S_1} \neq 0 $, $ g_{A_1} \neq 0 $, $ \xi_{S_1} \neq 0 $,
$ \xi_{A_1} \neq 0 $, $ n_{V_1} \neq 0 $) results in gradual
degradation of nonclassical properties of light,
i.e. the corresponding values of moments of intensity as well as of
the parameters characterizing squeezing successively increase.
Non-zero anti-Stokes coupling also introduces oscillations
to the spatial development of all the
quantities under discussion, in agreement with the fact that
the anti-Stokes interaction has tendency to conserve the initial
photon statistics.

\vspace{3mm}
\noindent
{\it Influence of $ \kappa_S $ on stimulated Stokes and
anti-Stokes processes in both waveguides}
\vspace{2mm}

In this case, nonclassical effects occur in the same
compound modes as for Brillouin processes discussed above,
i.e. negative moments of intensity and squeezing can occur in
compound modes ($ S_1,A_1 $), ($ S_1,V_1 $), ($ S_2,A_2 $),
($ S_2,V_2 $), ($ S_2,A_1 $), ($ S_2,V_1 $),
($ S_1,A_2 $), and ($ S_1,V_2 $).

Moments of intensity in mode ($ A_1,V_2 $) cannot be negative,
but values of moments for smaller $ z $ can be gradually
reduced when $ z $ increases, as is shown in Fig. 9.
\begin{figure}        
  \centerline{\hbox{\psfig{file=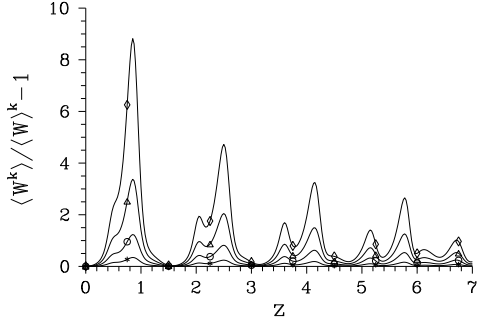,height=4.2cm}
     }}
  \vspace{3mm}
  \caption{Reduced moments of the integrated intensity
 $ \langle W^k(z)\rangle/\langle
 W(z)\rangle^k -1 $ for $ k=2 (\ast) $, $ k=3 (\circ) $,
 $ k=4 (\triangle)$, and $ k=5 (\diamond) $ for mode ($ A_1,V_2 $);
 $ g_{S_1} = 1 $, $ g_{A_1} = 2 $, $ \kappa_S = -6 $,
 $ g_{S_2} = 1 $, $ g_{A_2} = 2 $, $ \xi_{S_1} = -2i $,
 $ \xi_{A_1} = 2i $,
 $ n_{V_1} = 0.1 $, $ \xi_{S_2} = 2 $, $ n_{V_2} = 0.1 $,
 and the other parameters are zero.}
\end{figure}
\noindent

Non-zero values of $ \kappa_S $ introduce
oscillations to spatial development of discussed
quantities. Complex values of $ \kappa_S $ introduce
asymmetry into spatial development of the statistical quantities,
which affects mainly moments of intensity
(see Fig. 10 for moments ($ S_1,V_2 $) and ($ S_2,V_1 $)).
The reduction of initial noise in compound modes can be obtained,
including phonon modes $ V_1 $ and $ V_2 $.
\begin{figure}        
  \centerline{\hbox{\psfig{file=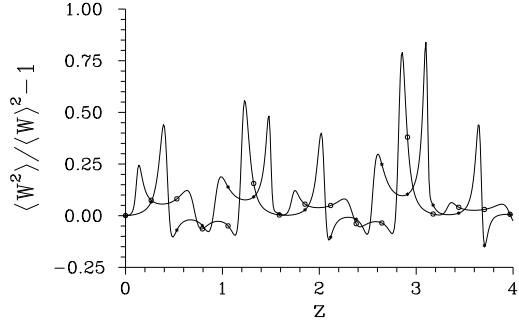,height=4.2cm}
     }}
  \vspace{3mm}
  \caption{The second reduced moments
 $ \langle W^2(z)\rangle/\langle W(z)\rangle^2 -1 $ for
 modes ($ S_1,V_2 $) $ (\ast) $ and ($ S_2,V_1 $) $ (\circ) $;
 $ g_{S_1} = 1 $, $ g_{A_1} = 2 $, $ \kappa_S = 6i $,
 $ g_{S_2} = 1 $, $ g_{A_2} = 2 $, $ \xi_{S_1} = -2i $,
 $ \xi_{A_1} = 2i $,
 $ n_{V_1} = 0.1 $, $ \xi_{S_2} = -2i $,
 $ \xi_{A_2} = 2i $, $ n_{V_2} = 0.1 $,
 and the other parameters are zero.}
\end{figure}

\vspace{3mm}
\noindent
{\it Influence of Stokes and anti-Stokes couplings on
 Raman processes in both waveguides}
\vspace{2mm}

The configuration from the previous section can be extended by
including anti-Stokes coupling ($ \kappa_A \neq 0 $).
Negative moments of intensity and squeezing occur in the
same compound modes as discussed in the previous section,
provided that the values of the anti-Stokes coupling constant $ \kappa_A $
are small. Greater values of $ \kappa_A $ provide greater values
of moments of intensity.
Fig. 11 shows the spatial development of
the second moment of intensity of mode ($ S_1,A_1 $), which
is negative only for smaller values of $ z $.
Greater values of $ \kappa_A $ lead to a complete loss of negative values
of moments
for compound modes, in which they can occur for
smaller values of $ \kappa_A $.
They also cause a successive
increase of values of variances and uncertainty with increasing $ z $
in the most of modes, but squeezing for greater $ z $ can be also
reached in some modes (e.g. ($ S_1,A_1 $), ($ S_2,A_2 $)).
\begin{figure}        
  \centerline{\hbox{\psfig{file=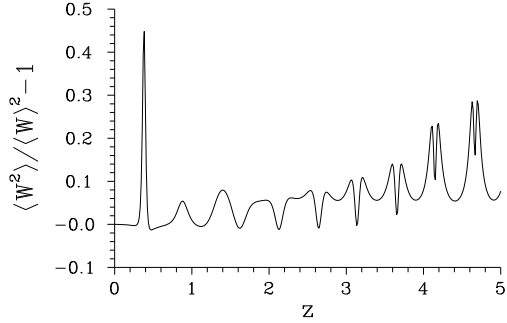,height=4.2cm}
     }}
  \vspace{3mm}
  \caption{The second reduced moment of the integrated intensity
 $ \langle W^2(z)\rangle/\langle W(z)\rangle^2 -1 $
 for mode ($ S_1,A_1 $);
  $ \kappa_A = 6i $
  and the other parameters are the same as in Fig. 10.}
\end{figure}
\noindent

Greater values of $ n_{V_1} $ and $ n_{V_2} $ preserve
negative moments of intensity and squeezing in compound
modes excluding phonon modes
(($ S_1,A_1 $), ($ S_2,A_2 $), ($ S_2,A_1 $), ($ S_1,A_2 $)).
They destroy nonclassical effects in compound modes involving
phonon modes $ V_1 $ and $ V_2 $.

Negative moments of intensity cannot be obtained in mode
($ V_1,V_2 $), but initial greater values of moments can be
substantially reduced.

\vspace{3mm}

If we compare the above results for couplers with
Brillouin and Raman processes, we can conclude that in
both the cases sub-Poissonian photon statistics and squeezing
of vacuum fluctuations can occur in both compound modes of one waveguide
(($ S_j,A_j $), ($ S_j,V_j $), $ j=1,2 $) and compound modes consisting
of single modes in different waveguides (($ S_1,A_2 $),
($ S_1,V_2 $)). The Stokes linear coupling supports nonclassical
properties of light, whereas the anti-Stokes linear coupling destroys
them. Both Stokes and anti-Stokes linear coupling constants
introduce oscillations to spatial development of the statistical
quantities under discussion.

\section{Conclusions}

We have investigated nonlinear optical couplers composed of two waveguides
operating by means of Brillouin or Raman processes, which mutually interact
through linear Stokes and anti-Stokes couplings. We have solved
the model, assuming classical strong laser pump modes, analytically
for special cases and in short-length approximation as well.
General solutions have been obtained numerically.
We have investigated the statistical properties of single
and compound modes, in particular we have obtained
photon number distributions, reduced moments
of the integrated intensity, quadrature variances, principal
squeeze variance and uncertainty product. It has
been shown that nonclassical light can be generated
only in the compound modes. Light exhibiting
sub-Poissonian photon
statistics and squeezing of vacuum fluctuations can be
obtained in both compound modes of one waveguide (($ S_j,A_j $),
($ S_j,V_j $), $ j=1,2 $) and compound modes composed of single modes in
different waveguides (($ S_1,A_2 $),($ S_1,V_2 $)).
The linear Stokes coupling supports nonclassical light
generation, whereas the linear anti-Stokes coupling leads
to its degradation. Both Stokes and anti-Stokes couplings
introduce oscillations into spatial development of the statistical
quantities. The above conclusions
are valid for both Brillouin and Raman processes. However,
in Raman processes, increasing values of the mean phonon
numbers degrade nonclassical light, especially
in compound modes involving phonon modes.

\section{Acknowledgments}

This work was supported by the Complex Grant VS96028 of Czech Ministry
of Eduation and by the Grant 202/96/0421 of Czech Grant Agency.

\end{document}